\documentclass[twocolumn,twoside]{IEEEtran}

\def\fighome {figures/}

\usepackage[cmex10]{amsmath}
\usepackage{epsfig,epsf,psfrag,amssymb,amsfonts,latexsym,slashbox,graphicx,bm,cite,xcolor}
\usepackage[caption=false,font=footnotesize]{subfig}%
\usepackage{fixltx2e}
\usepackage{array}
\usepackage{cases}
\usepackage{verbatim}
\usepackage[mathscr]{eucal}
\usepackage{footmisc}

\usepackage[T1]{fontenc}
\usepackage[latin1]{inputenc}
\usepackage{float}
\floatstyle{ruled}
\newfloat{algorithm}{tbp}{loa}
\floatname{algorithm}{Algorithm}

\DeclareMathOperator{\unif}{Unif}

\def\leqst{{\,\,\overset{st}{\leq}\,\,}}

\def\tha{{\mbox{\tiny th}}}

 \def\0{{\bf 0}}

\def\viz{{viz.,\ \/}}
\def\ie{{i.e.,\ \/}}

\def\nn{\nonumber}

\def\qed{\hfill\hbox{${\vcenter{\vbox{
    \hrule height 0.4pt\hbox{\vrule width 0.4pt height 6pt
    \kern5pt\vrule width 0.4pt}\hrule height 0.4pt}}}$}}

\definecolor{myred}{rgb}{0.3,0.0,0.7}
\definecolor{dkg}{rgb}{0.1,0.7,0.2}
\definecolor{dkb}{rgb}{0.0,0.2,0.8}

\newcommand{\Gmsc}{\mathscr{G}}

\def\bfg{{\mathbf g}}

\def\bfX{{\mathbf X}}

\def\bfZ{{\mathbf Z}}    

\def\mubf{\hbox{\boldmath$\mu$\unboldmath}}

\def\Ac{{\cal A}}

\def\Cc{{\cal C}}

\def\Hc{{\cal H}}

\def\Xc{{\cal X}}

\def\Ebb{{\mathbb E}}

\def\Nbb{{\mathbb N}}

\def\Pbb{{\mathbb P}}

\newcommand{\bprf}{\begin{myproof}}
\newcommand{\eprf}{\end{myproof}}
\newcommand{\bp}{\begin{psfrags}}
\newcommand{\ep}{\end{psfrags}}
\newcommand{\bl}{\begin{lemma}}
\newcommand{\el}{\end{lemma}}
\newcommand{\bt}{\begin{theorem}}
\newcommand{\et}{\end{theorem}}
\newcommand{\bc}{\begin{center}}
\newcommand{\ec}{\end{center}}
\newcommand{\bi}{\begin{itemize}}
\newcommand{\ei}{\end{itemize}}
\newcommand{\ben}{\begin{enumerate}}
\newcommand{\een}{\end{enumerate}}
\newcommand{\bd}{\begin{definition}}
\newcommand{\ed}{\end{definition}}
\def\beq{\begin{equation}}
\def\eeq{\end{equation}\noindent}
\def\beqn{\begin{eqnarray}}
\def\eeqn{\end{eqnarray} \noindent}
\def\beqnn{  \begin{eqnarray*}}
\def\eeqnn{\end{eqnarray*}  \noindent}
\def\bcase{  \begin{numcases}}
\def\ecase{\end{numcases}   \noindent}
\def\bsbcase{  \begin{subnumcases}}
\def\esbcase{\end{subnumcases}   \noindent}

\def\defeq{{:=}}

\newtheorem{theorem}{Theorem}

\newtheorem{lemma}{Lemma}
\newtheorem{definition}{Definition}

\newtheorem{proposition}{Proposition}

\newenvironment{myproof}{\noindent{\em Proof:} \hspace*{1em}}{
    \hspace*{\fill} $\Box$ }
\newenvironment{proof_of}[1]{\noindent {\em Proof of #1:}
    \hspace*{1em} }{\hspace*{\fill} $\Box$ }

\def\psfancypar#1#2{\begingroup\def\par{\endgraf\endgroup\lineskiplimit=0pt}
               \setbox2=\hbox{\large\sc #2}
               \newdimen\tmpht \tmpht \ht2 \advance\tmpht by \baselineskip
               \font\hhuge=Times-Bold at \tmpht
               \setbox1=\hbox{{\hhuge #1}}
               \count7=\tmpht \count8=\ht1
               \divide\count8 by 1000 \divide\count7 by \count8 
               \tmpht=.001\tmpht\multiply\tmpht by \count7 
               \font\hhuge=Times-Bold at \tmpht
               \setbox1=\hbox{{\hhuge #1}}
               \noindent
                \hangindent1.05\wd1
               \hangafter=-2 {\hskip-\hangindent
               \lower1\ht1\hbox{\raise1.0\ht2\copy1}%
                \kern-0\wd1}\copy2\lineskiplimit=-1000pt}

\def\Kout{\setbox1=\hbox{\Huge\bf K}\hbox to
1.05\wd1{\hspace{.05\wd1}
}}
\def\Sout{\setbox1=\hbox{\Huge\bf S}\hbox to 1.05\wd1{\hspace{.05\wd1}
}}

\def\unknownalloc{{\rho^{\mbox{\tiny EST}}}}
\def\randalloc{{\rho^{\mbox{\tiny RAND}}}}
\def\policy{{\rho}}

\def\centpi{{\rho^{\mbox{\tiny CENT}}}}
\def\cent{{\rho^{\mbox{\tiny CENT}}}}
\def\singleuser{{\rho^1}}

\def\statbf{\bfg}

\def\unif{\mbox{Unif}}
\def\iscollide{{\zeta}}

\def\curchannel{{Curr\_Sel}}
\def\currank{{Curr\_Rank}}
\def\est{{\mbox{\tiny 
EST}}}

\begin{document}

\title{Distributed Algorithms for Learning and \\Cognitive Medium  Access with Logarithmic Regret} 

\author{Animashree Anandkumar$^\dagger$,
\thanks{$^\dagger$\scriptsize Corresponding
author.}~\IEEEmembership{Member,~IEEE}, Nithin Michael,~\IEEEmembership{Student Member,~IEEE}, \\  Ao Kevin Tang,~\IEEEmembership{Member,~IEEE}, and Ananthram
Swami,~\IEEEmembership{Fellow,~IEEE}.
\thanks{\scriptsize
A. Anandkumar is  with the School of Electrical Engineering and
Computer Science, MIT, Cambridge, MA 02139, USA. Email: {\tt
animakum@mit.edu}  }
 \thanks{\scriptsize  N. Michael and A.K. Tang are with the School
 of Electrical and Computer Engineering, Cornell University, Ithaca,
 NY 14853, USA.  Email:
{\tt{nm373@,atang@ece.}cornell.edu}} \thanks{\scriptsize A. Swami is
with the Army Research Laboratory, Adelphi, MD 20783, USA. E-mail:
{\tt a.swami@ieee.org}.}
\thanks{\scriptsize The first author is  supported by MURI through
 AFOSR Grant FA9550-06-1-0324. The second and the
 third authors are supported in part through NSF grant CCF-0835706.}
\thanks{\scriptsize
Parts of this paper were presented at
\cite{Anandkumar&etal:INFOCOM10}.}
}

\maketitle

\begin{abstract}The problem of distributed learning and channel access  is considered   in a cognitive network  with multiple secondary users. The availability statistics of the channels are initially unknown to the secondary users and are estimated using sensing 
decisions. There is no explicit information exchange or prior 
agreement  among the secondary users. We propose policies  for distributed learning and access which achieve order-optimal  cognitive system throughput (number of successful secondary transmissions) under self play, i.e., when implemented at all the secondary users. Equivalently, our 
policies minimize the regret in distributed learning and access. We first consider the scenario when the number of secondary users is known to the policy, and prove  that the total regret is logarithmic in the number of transmission slots.   Our distributed learning and access policy achieves order-optimal   regret  by comparing  to an asymptotic  lower bound for   regret  under any uniformly-good learning and access policy. We then consider the case when    the number of secondary users is fixed but unknown, and is estimated through feedback.  We propose a policy in this scenario whose  asymptotic sum regret which grows  slightly faster than logarithmic in the number of transmission slots.
\end{abstract}
 

\section{Introduction}

There has been extensive research on cognitive radio network in the 
past decade  to resolve many  challenges not  encountered previously in traditional communication networks (see  \cite{Zhao&Sadler:07SPM}). One of the main challenges is to achieve   coexistence of heterogeneous users accessing the same part of the spectrum.
In a typical cognitive network, there are two classes of transmitting users, \viz the 
primary users who have priority in accessing the spectrum and the 
secondary users who   opportunistically transmit when 
the primary user is idle. The secondary users are {\em cognitive} and can sense the spectrum to   detect the presence of a primary transmission. However, due to resource and 
hardware constraints, they can  sense  only a part of the spectrum 
at any given time. 

\begin{figure}[t]\bc
\bp \psfrag{p}[r]{\small Primary User}
 \psfrag{secondary}[l]{\small Secondary User}
\includegraphics[width=2.3in]{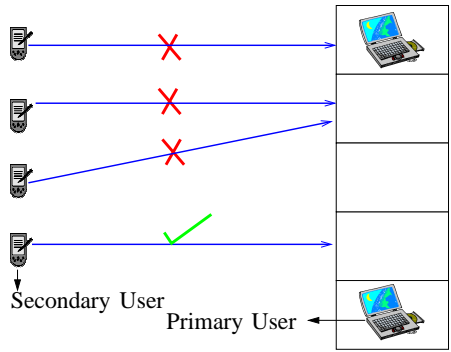}\ep\ec
\caption{Cognitive radio network with $U=4$ secondary users and
$C=5$ channels.   A secondary user is not allowed to transmit if the
accessed channel is occupied by a primary user. If more than one
secondary user transmits in the same free channel, then all the
transmissions are unsuccessful.}\label{fig:cognitive}\end{figure}

We consider  a slotted cognitive system where each secondary user can  sense and access only one orthogonal channel in each transmission slot  (see Fig.\ref{fig:cognitive}).
Under sensing constraints, it is thus   beneficial for the secondary users to select channels with 
higher mean availability, \ie channels which are less likely to be occupied by the  primary users.  However, in practice, the channel availability statistics are a priori unknown  to the secondary users. 

Since the secondary users are required to sense the medium before transmission, can these sensing decisions   be used to  {\em learn} the  channel availability statistics? If so,   using these   estimated channel availabilities, can we design channel access rules which maximize the transmission throughput? Designing provably efficient algorithms to accomplish the above goals forms the focus of our paper.  Such  algorithms need to be efficient, both in terms of learning and channel access.

For any learning algorithm, there are two important performance criteria: convergence and {\em regret} bounds \cite{Cesa:book}.  In the above context, we require the estimates  to converge to the correct channel availability statistics as the number of available sensing decisions  goes to infinity. A stronger criterion is the regret of a learning algorithm, which measures the speed of convergence. 
In our context, the regret is the loss in secondary throughput due to learning     compared with knowing the channel  statistics perfectly.  Hence, it is desirable for the learning algorithms to have small regret.
The regret is a finer measure of performance of a learning algorithm than the time-averaged throughput since a sub-linear regret (with respect to time) implies optimal average throughput.

Additionally, we consider a distributed  framework where there is no information exchange or prior agreement among the secondary users.  This introduces additional  challenges: it results in   loss of throughput due to   collisions among the secondary users, and  there is  now  competition among  the secondary users since they all tend to access channels with higher availabilities. It is imperative for the channel access policies to overcome the above challenges.  Hence, a distributed learning and access policy experiences  regret both  due to learning of the
unknown channel availabilities as well as due to collisions under
distributed access.

\subsection{Our Contributions}\label{sec:contributions}

The main contributions of this paper are two fold. First, we propose two distributed learning and access policies for multiple secondary users in a cognitive network. Second, we provide performance guarantees for these policies in terms of regret.  Overall, we prove that one of  our proposed algorithms achieves order-optimal regret and the other achieves nearly order-optimal regret, where the order is in terms of the number of transmission slots.  


The  first policy we propose   assumes that the total number of secondary users in the system is known while our second policy relaxes this requirement. Our second policy also incorporates estimation of the number of secondary users, in addition to learning of the channel availabilities and designing distributed access rules. We provide bounds on total regret experienced by the secondary users under self play, i.e., when implemented at all the secondary users. For the first policy, we prove that the regret is logarithmic, i.e., $O(\log n)$ where $n$ in the number of transmission slots. For the second policy, the regret  grows slightly faster than logarithmic, \ie $O(f(n) \log n)$, where we can choose any function $f(n)$ satisfying $  f(n) \to \infty$, as $n \to \infty$. Hence, we provide performance guarantees for the proposed distributed learning and access policies. 

A lower bound on regret under   any   uniformly-good distributed learning policy has been derived in \cite{Liu&Zhao:09Arxiv}, which is also logarithmic in the number of transmission slots. Thus, our first policy (which requires  knowledge of the number of secondary users) achieves order-optimal regret. The effects of the      
number of secondary users and the   number of channels on regret are 
also explicitly characterized and verified via simulations. 

To the best of our knowledge, the {\em exploration-exploitation} tradeoff for learning, combined with the  {\em cooperation-competition}  tradeoffs among multiple  users for distributed medium access  have not been sufficiently examined in the
literature before (see Section~\ref{sec:related} for a discussion).  Our analysis in this paper  provides important engineering insights towards dealing with learning, 
competition,  and cooperation in  practical cognitive systems. 
 

{\em Remark:} We note some of the shortcomings of our approach.
The i.i.d. model\footnote{By i.i.d. primary transmission model, we
do not mean the presence of a single primary user, but rather, this model is used to capture the overall statistical behavior of all the primary users in the system.} for primary
transmissions is indeed   idealistic and in practice,  a Markovian
model may be more appropriate \cite{Konrad:03,Geirhofer&Tong&Sadler:08JSAC}.
However, the i.i.d. model is a good approximation if the time 
slots for transmissions are  long and/or the primary 
traffic is highly bursty.     Moreover, the i.i.d. model is not crucial towards deriving regret bounds  for our proposed schemes. Extensions of the classical 
multi-armed bandit problem to a Markovian  model  are 
considered  in \cite{Anantharam&etal2:87TranAC}. In principle, our results 
on distributed learning and access can be similarly extended to 
a Markovian channel model  but this entails more complex estimators 
and rules for evaluating the exploration-exploitation tradeoffs of 
different channels and is a topic of interest for future 
investigation.

\subsection{Related Work}\label{sec:related}

Several results on the multi-armed bandit problem will be used and 
generalized to study our problem. Detailed discussion on multi-armed 
bandits can be found in 
\cite{Lai&Robbins:85AAM,Anantharam&etal:87TranAC,Agrawal:95AAP,Auer&etal:02ML}. 
Cognitive medium access is a topic of extensive research; see \cite{Zhao:bookchapter} for an overview. The connection between cognitive medium access and the multi-armed bandit problem is explored in 
\cite{Liu&Zhao:08IT}, where a restless bandit formulation is employed. Under this formulation,    indexability   is established, the Whittle's index for channel selection is   obtained in closed-form, and the equivalence between the  myopic policy and the Whittle's index is established. However, this work assumes known channel availability statistics   and does not consider competing secondary  users.  The work in \cite{Liu&etal:08ICC}   considers allocation of two users to two 
channels    under Markovian channel model   using a partially 
observable Markov decision process (POMDP) framework. The use of  collision feedback information for learning, and spatial heterogeneity in spectrum opportunities were investigated. However, the 
difference from our work is that \cite{Liu&etal:08ICC} assumes that 
the availability statistics (transition probabilities) of the 
channels are known to the secondary users  while we consider 
learning of unknown channel statistics.  The works in 
\cite{Fu&Schaar:09TranVT,Gang&etal:08crown} consider centralized 
access schemes in contrast to distributed access here, 
\cite{Liu&etal:08ICWI} considers access through information 
exchange and studies the optimal choice of the amount of information to be exchanged given the cost of negotiation.   \cite{Li:09SMC} considers access under 
$Q$-learning for  two  users  and two channels where  users can 
sense both the channels simultaneously. 
The work in  \cite{Maskery&etal:chapter} discusses a game-theoretic approach to cognitive medium access. In \cite{Kleinberg&etal:09STOC},  learning  in congestion games 
through multiplicative updates is considered and convergence to weakly-stable equilibria (which reduces to the pure Nash equilibrium for almost all games) is proven. However, the work assumes  fixed costs (or equivalently rewards) in contrast to random rewards here,  and that the players can fully observe the actions of other players.

Recently, the work in \cite{Gai&etal:dyspan10} considers combinatorial bandits, where a more general model of different (unknown)  channel availabilities is assumed for different secondary users, and a  matching algorithm is proposed for jointly allocating users to channels. The algorithm is guaranteed to have logarithmic regret with respect to number of transmission slots and polynomial storage requirements. A decentralized implementation of the proposed algorithm is proposed but it still requires information exchange and coordination among the users. In contrast, we propose algorithms which removes this requirement albeit in a more restrictive setting. 


In our recent work \cite{Anandkumar&etal:INFOCOM10}, we first 
formulated the problem of decentralized learning and access for 
multiple secondary users. We considered two scenarios:  one where there is initial common information among the secondary users in the form of pre-allocated ranks, and the other where no such information is available.  In this paper, we analyze the distributed 
policy in detail and prove that it has logarithmic regret.  In addition, we also consider the case 
when the  number of secondary users   is unknown, and provide bounds on regret in this scenario.

Recently,   Liu and Zhao \cite{Liu&Zhao:09Arxiv} proposed a family of distributed learning and access policies known as time-division fair share (TDFS), and proved logarithmic regret for these policies. They established a lower bound  on the growth rate of system regret  for a general class of uniformly-good decentralized polices.  The TDFS policies in \cite{Liu&Zhao:09Arxiv} can incorporate any order-optimal single-player policy while our work here is based on  the single-user policy proposed in  \cite{Auer&etal:02ML}. Another difference is that in \cite{Liu&Zhao:09Arxiv}, the users  orthogonalize via settling at different offsets in their time-sharing schedule, while in our work here, users orthogonalize into different channels. Moreover, the TDFS policies ensure   that each player achieves the same time-average reward while our policies here achieve probabilistic fairness, in the sense that the policies do not discriminate between different users. In \cite{Liu&etal:10Asilomar}, the TDFS policies are extended to incorporate  imperfect sensing.

\subsubsection*{Organization \& Suggested Reading} Section
\ref{sec:model} deals with the system model, Section
\ref{sec:special} deals with the special case of single secondary
user and of multiple users with centralized access which can be
directly solved using the classical results on multi-armed 
bandits. In Section \ref{sec:results}, we propose  distributed learning and access policy with provably logarithmic regret when the number of secondary users is known. Section~\ref{sec:unknown} 
considers the scenario when the 
number of secondary users is unknown.   Section~\ref{sec:lowerbound} provides a lower bound 
for distributed learning. Section~\ref{sec:num} has 
simulation results for the proposed schemes and Section 
\ref{sec:conclusion} concludes the paper. Most of the proofs are 
found in the Appendix.

Since Section   \ref{sec:special} mostly deals with a recap of the
classical results on multi-armed bandits, we suggest that an
experienced reader directly jump to Section~\ref{sec:results} for
the  main results of this paper.

\section{System Model \& Formulation}\label{sec:model}

\subsubsection*{Notation}

For any two functions $f(n),g(n)$, $f(n) = O(g(n))$ if there exists a constant $c$ such that $f(n) \leq c g(n)$ for all $n \geq n_0$ for a fixed $n_0\in \Nbb$. Similarly, $f(n) = \Omega(g(n))$ if there exists a constant $c'$ such that $f(n) \geq c'g(n)$  for all $n \geq n_0$ for a fixed $n_0\in \Nbb$, and $f(n) = \Theta(g(n))$ if $f(n)= \Omega(g(n))$ and $f(n) = O(g(n))$. Also, $f(n) = o(g(n))$ when $f(n)/g(n) \to 0$ and $f(n) = \omega(g(n))$ when $f(n)/ g(n) \to \infty$ as $n \to \infty$.

We refer to the $U$ highest entries in a vector $\mubf$ as the $U$-best
channels and the rest  as the $U$-worst channels. Let
$\sigma(T;\mubf)$   denote    the index of  the $T^{\tha}$ highest
entry in $\mubf$. Alternatively, we abbreviate
$T^*\defeq\sigma(T;\mubf)$ for ease of notation. With abuse of
notation, let $D(\mu_1,\mu_2)\defeq D(B(\mu_1);B(\mu_2))$ be the
Kullback-Leibler distance between the
Bernoulli distributions $B(\mu_1)$ and  $B(\mu_2)$
\cite{Cover&Thomas:book} and let $\Delta(1,2)\defeq \mu_1- \mu_2$.

\subsection{Sensing \& Channel Models}
Let $U\geq 1$ be the number of secondary users\footnote{A user
refers to a secondary user unless otherwise mentioned.}    and $C\geq U$
be  the number\footnote{When $U\geq C$, learning availability statistics is less crucial, since all channels need to be accessed to avoid collisions. In this case, design of medium access is more crucial.} of orthogonal channels available for slotted
transmissions with a fixed slot width. In each channel $i$ and slot
$k$, the primary user transmits i.i.d. with probability $1-\mu_i>0$. In other words, let $W_i(k)$ denote the indicator variable if the channel is
free \bcase{W_i(k)=} 0, & channel $i$ occupied in slot $k$\nn\\ 1, &
o.w,\nn \ecase and we assume that $W_i(k) \overset{i.i.d.}{\sim} B(\mu_i)$.

The mean availability vector $\mubf$ consists of mean availabilities $\mu_i$ of all channels, i.e.,  is  $\mubf\defeq[\mu_1, \ldots, \mu_C]$, where all $\mu_i \in (0,1)$ and are distinct. $\mubf$ is initially unknown to all the secondary users  and  is learnt {\em independently} over time using the past   sensing decisions without  any information exchange among the users. We assume that sensing for primary transmissions is perfect at all the users.

Let  $T_{i,j}(k)$ denote the number of
slots where channel $i$ is sensed in $k$ slots by user $j$  (not necessarily
being the sole occupant of that channel).
The sensing variables are obtained as follows: at the beginning of
each slot $k$, each secondary user $j \in U$ selects exactly  one
channel $i \in C$ for   sensing, and hence, obtains the
value of $W_i(k)$, indicating if the channel is
free. User $j$ then records all the sensing decisions of
each channel $i$ in a vector $\bfX_{i,j}^k\defeq[X_{i,j}(1),\ldots,
X_{i,j}(T_{i,j}(k))]^T$. Hence, $\cup_{i=1}^C\bfX_{i,j}^k$
is the collection of sensed decisions
for user $j$ in $k$ slots for all the $C$ channels.




We assume  the collision model under which if two or more users
transmit in the same channel then none of the transmissions go
through.    At the end of each slot $k$, each user $j$ receives
acknowledgement $Z_j(k)$ on whether its transmission in the
$k^{\tha}$ slot was received. Hence, in general, any policy employed
by user $j$ in the $(k+1)$-th slot, given by $\policy( \cup_{i=1}^C\bfX_{i,j}^k,\bfZ_j^k)$ is based on all the previous sensing and  feedback results.

\subsection{Regret of a Policy}
 
Under the above model, we are interested in designing policies
$\policy$ which   maximize the expected number of successful
transmissions of the secondary users subject to the non-interference
constraint for the primary users. Let $S(n;\mubf,U, \policy)$ be the
expected total number of successful transmissions after $n$ slots
under $U$ number of secondary  users and policy $\policy$.

In the ideal scenario where the  availability statistics $\mubf$
 are known a priori and a central agent orthogonally
 allocates the secondary users to the $U$-best channels, the expected number of successful transmissions after $n$ slots is given by \beq S^*(n;\mubf,U) \defeq n \sum_{j
=1}^U \mu(j^*),\label{eqn:opt_reward}\eeq where $j^*$ is the
$j^{\tha}$-highest entry in $\mubf$.

It is clear that  $S^*(n;\mubf,U)
> S(n;\mubf,U,\policy)$ for any policy $\policy$ and finite $n$. We are interested in
minimizing the {\em regret} in learning and access, given by\beq
R(n;\mubf,U,\policy)\defeq S^*(n;\mubf,U) -
S(n;\mubf,U,\policy)>0.\label{eqn:regret_def}\eeq 
We are interested in minimizing regret under any given $\mubf\in
(0,1)^C$ with distinct elements.

By incorporating the collision channel model assumption with no
avoidance mechanisms\footnote{The effect of employing  CSMA-CA is
not considered here although it can be shown that it reduces the
regret and hence, the bounds we derive are applicable.}, the expected throughput under policy $\policy$ is given by \[ S(n;\mubf,U,\policy) = \sum_{i=1}^C \sum_{j=1}^U \mu(i)\Ebb[V_{i,j}(n)],\]  where
$V_{i,j}(n)$ is the   number of times in $n$ slots where user
$j$ is the sole user to sense channel $i$. Hence, the regret
in \eqref{eqn:regret_def} simplifies as     \beq R(n;\policy)=
\sum_{k =1}^U n\mu(k^*)-\sum_{i=1}^C \sum_{j=1}^U \mu(i)
\Ebb[V_{i,j}(n)].\label{eqn:regret_collision}\eeq

\section{Special Cases From Known Results}\label{sec:special}

We  recap the bounds for the regret under the special cases of a
single secondary user $(U=1)$ and multiple users with centralized
learning and  access by appealing to the classical results on
the multi-armed bandit process
\cite{Lai&Robbins:85AAM,Anantharam&etal:87TranAC,Agrawal:95AAP}.

\subsection{Single Secondary User $(U=1)$}

\begin{algorithm} {\bf Input:} $\{\bar{X}_{i}(n)\}_{i=1,\ldots,C}:$ Sample-mean availabilities   
 after $n$ rounds, $g(i;n)$: statistic based on
$\bar{X}_{i,j}(n)$,  \\ $\sigma(T;\statbf(n))$: index of $T^{\tha}$
highest entry in $\statbf(n)$.\\ Init:  Sense  in each channel
once, $n \leftarrow C$ \\ Loop: $n\leftarrow n+1$  \\ 
$\curchannel\leftarrow$ channel corresponding to highest entry in $\statbf(n) $ for
sensing. If free, transmit.      \caption{Single User Policy $\singleuser(\statbf(n))$ in  \cite{Agrawal:95AAP}.}
 \label{algo:singleuser} 
 \end{algorithm}

%
%
%

When there is only one secondary user, the problem of finding policies with minimum regret reduces to that of a multi-armed bandit process. Lai and Robbins \cite{Lai&Robbins:85AAM} first analyzed
schemes for multi-armed bandits with asymptotic logarithmic regret
based on the upper confidence bounds on the unknown channel
availabilities. Since then, simpler schemes have been proposed in
\cite{Agrawal:95AAP,Auer&etal:02ML} which compute a statistic or an index for
each arm (channel), henceforth referred to as the {\em
$g$-statistic}, based only on its sample mean and the number of
slots where the particular arm is sensed.  The arm with the
highest index is selected in each slot in these works. We summarize the policy in Algorithm \ref{algo:singleuser} and denote it $\singleuser(\statbf(n))$, where $\statbf(n)$ is the vector of scores assigned to the channels after $n$ transmission slots.


The sample-mean based policy  in \cite[Thm. 1]{Auer&etal:02ML} proposes an index for each channel  $i$ and user $j$ at time $n$ is  given by\beq g_j^{\mbox{\tiny
MEAN}}(i;n)\defeq \bar{X}_{i,j}(T_{i,j}(n)) + \sqrt{\frac{2 \log
n}{T_{i,j}(n)}},\label{eqn:sample_mean}\eeq where $T_{i,j}(n)$ is the number of slots
 where user $j$ selects channel $i$ for sensing
and \beq
\bar{X}_{i,j}(T_{i,j}(n))\defeq\sum_{k=1}^{T_{i,j}(n)}\frac{
X_{i,j}(k)}{T_{i,j}(n)}\nonumber \eeq is the sample-mean
availability of channel $i$, as sensed by user $j$. 

The   statistic in \eqref{eqn:sample_mean}  captures the {\em exploration-exploitation} tradeoff  between   sensing the channel with the best predicted availability to maximize  immediate throughput and sensing different  channels to obtain improved estimates of their availabilities. 
The sample-mean term in \eqref{eqn:sample_mean} corresponds to exploitation while the other term involving $T_{i,j}(n)$ corresponds to exploration since it penalizes channels which are not sensed often.  The normalization of the exploration term  with $\log n$ in \eqref{eqn:sample_mean} implies that the term is significant when $T_{i,j}(n)$  is much smaller than $\log n$. On the other hand, if all the channels
are sensed $\Theta(\log n)$ number of times, the exploration terms
become unimportant in the $g$-statistics of the channels and the exploitation term dominates, thereby, favoring sensing of the channel with the highest sample mean.

The regret based on the
above statistic in \eqref{eqn:sample_mean} is logarithmic   for any finite number of
slots $n$ but does not have the optimal scaling constant.
The sample-mean based statistic in \cite[Example 5.7]{Agrawal:95AAP}  leads to the optimal scaling constant for regret and is given by \beq\label{eqn:gopt} g_j^{\mbox{\tiny OPT}}(i;n)\defeq
\bar{X}_{i,j}(T_{i,j}(n)) + \min\left[\sqrt{\frac{\log n}{2
T_{i,j}(n)}},1\right].\eeq In this
paper, we design policies based on the $g^{\mbox{\tiny
MEAN}}$ statistic since it is simpler
to analyze than the $g^{\mbox{\tiny
OPT}}$ statistic.

We now recap the results which show logarithmic regret in learning
the best channel. In this context, we define  {\em uniformly good} policies
$\policy$ \cite{Lai&Robbins:85AAM} as those with regret\beq R(n;\mubf,U,
\policy) = o(n^\alpha), \quad \forall \alpha >0,
\mubf\in(0,1)^C.\label{eqn:unif_good}\eeq

\bt[Logarithmic Regret for $U=1$
\cite{Agrawal:95AAP,Auer&etal:02ML}]\label{thm:lowerbnd_singleuser}For any
uniformly good policy $\policy$ satisfying \eqref{eqn:unif_good},
the expected time spent in any suboptimal  channel $i \neq 1^*$ satisfies
\beq\label{eqn:lowerbnd_tij_singleuser}\lim_{n\to
\infty}\Pbb\left[T_{i,1}(n) \geq \frac{(1-\epsilon)\log n}{ D(
\mu_i,\mu_{1^*})};\mubf\right]=1,\eeq where $1^*$ is the channel with the best availability. Hence, the regret satisfies
\beq \liminf_{n \to \infty} \frac{R(n;\mubf,1,\policy)}{\log n} \geq
\sum_{i \in 1\mbox{\scriptsize-worst}} \frac{\Delta(1^*,i)}{D(
\mu_i,\mu_{1^*})}.\label{eqn:lowerbndregret_singleuser}\eeq The regret under the $ g^{\mbox{\tiny OPT}}$
statistic in \eqref{eqn:gopt}  achieves the above bound.\beq \lim_{n
\to \infty} \frac{R(n;\mubf,1,\singleuser(\bfg_j^{\mbox{\tiny
OPT}}))}{\log n} = \sum_{i \in 1\mbox{\scriptsize-worst}}
\frac{\Delta(1^*,i)}{D(\mu_i,\mu_{1^*})}.\label{eqn:lowerbndregret_singleuser_opt}\eeq
The regret under  $g^{\mbox{\tiny MEAN}}$ statistic in \eqref{eqn:geomean} satisfies
\beq R(n;\mubf,1,
\singleuser(\bfg_j^{\mbox{\tiny MEAN}}))\leq \sum_{i \neq 1^*}
\Delta(1^*, i)\left[\frac{ 8 \log n}{\Delta(j^*,i)^2} +
1+\frac{\pi^2}{3}\right].\!\!\! \!\nonumber \eeq   \et
 
\subsection{Centralized Learning \&  Access for Multiple Users}

We now consider multiple secondary users under centralized access policies where there is joint learning and access by a central agent on behalf of all the $U$ users.  Here, to minimize the sum
regret, the centralized policy  allocates the $U$ users to orthogonal channels to avoid collisions. Let  $ \centpi(\Xc^k)$, with $\Xc^k := \cup_{j=1}^U \cup_{i=1}^C\bfX_{i,j}^k$, denote a centralized policy based on  the sensing  variables of all the users.  The policy under centralized learning is a simple generalization of the single-user policy and is given in Algorithm~\ref{algo:cent}. We now recap the results of \cite{Anantharam&etal:87TranAC}.



\bt[Regret Under Centralized Policy $ \centpi$
\cite{Anantharam&etal:87TranAC}]\label{thm:lowerbnd_cent}For any
uniformly good centralized policy $\centpi$ satisfying
\eqref{eqn:unif_good}, the expected times spent in a $U$-worst
channel $i$ satisfies  \beq\label{eqn:lowerbnd_ti}\lim_{n\to
\infty}\Pbb\left[\sum_{j=1}^U T_{i,j}(n) \geq \frac{(1-\epsilon)\log
n}{ D(\mu_i,\mu_{U^*})};\mubf\right]=1,\eeq where $U^*$ is the channel with the $U^{\tha}$ best availability. Hence, the regret
satisfies  \beq \liminf_{n \to \infty}
\frac{R(n;\mubf,1,\centpi)}{\log n} \geq \sum_{i \in
U\mbox{\scriptsize-worst}} \frac{\Delta(U^*,i)
}{D(\mu_i,\mu_{U^*})}.\label{eqn:lowerbndregret_central}\eeq The
scheme in Algorithm~\ref{algo:cent} based on $g^{\mbox{\tiny OPT}}$
  achieves the above bound.  \beq \lim_{n \to \infty}
\frac{R(n;\mubf,1,\centpi(\bfg^{\mbox{\tiny OPT}})}{\log n} =
\sum_{i \in U\mbox{\scriptsize-worst}} \frac{\Delta(U^*,i)
}{D(\mu_i,\mu_{U^*})}.\label{eqn:regret_cent_opt}\eeq The scheme in
Algorithm~\ref{algo:cent} based on the  $g^{\mbox{\tiny MEAN}}$ satisfies for any   $n>0$, \begin{align}\nn &R(n;\mubf,U, \cent(\statbf^{\mbox{\tiny
MEAN}}))\\&\leq
 \sum_{m =1}^U \sum_{i\in U\mbox{\scriptsize-worst}}
\sum_{k=1}^U \frac{\Delta(m^*,i) }{U}\left[\frac{ 8 \log
n}{\Delta(m^*,i)^2} +
1+\frac{\pi^2}{3}\right].\label{eqn:regret_cent_mean}\end{align}   \et

\bprf  See Appendix~\ref{proof:lowerbnd_cent}.\eprf

\begin{algorithm} {\bf Input:} $\Xc^n:=\cup_{j=1}^U\cup_{i=1}^C\bfX_{i,j}^n:$ Channel availability
 after $n$ slots,  $\statbf(n)$: statistic based on
$\Xc^n$,  \\ $\sigma(T;\statbf(n))$: index of
$T^{\tha}$ highest entry in $\statbf(n)$.\\ Init:  Sense  in
each channel once, $n \leftarrow C$ \\ Loop: $n\leftarrow n+1$
\\ $\curchannel\leftarrow$   channels with  $U$-best entries in
$\statbf(n) $. If free, transmit.   \caption{Centralized Learning Policy $\cent$ in  \cite{Anantharam&etal:87TranAC}.}
 \label{algo:cent}
 \end{algorithm}
 
\section{Main Results}\label{sec:results}

Armed with the classical results on multi-armed bandits, we now design distributed learning and allocation policies. 

\subsection{Preliminaries: Bounds on Regret}

We first provide simple bounds on the  regret   in
\eqref{eqn:regret_collision} for any distributed learning and access policy $\policy$.

\begin{proposition}[Lower and Upper Bounds on Regret]The regret under any distributed policy $\policy$ satisfies
\label{prop:bounds}
 \begin{align}   R(n;\policy)\geq& \sum_{j=1}^U
\sum_{i\in U\mbox{\scriptsize-worst}} \Delta(U^*,i)\Ebb[
T_{i,j}(n)],\label{eqn:regret_Mlowerbnd}\\
   R(n;\policy)\leq&     \mu(1^*)\!\!\left[\sum_{j=1}^U
  \sum_{i\in U \mbox{\scriptsize-worst}} \!\!\!\!\!\!  \Ebb[ T_{i,j}(n)] +
\Ebb[ M(n)]\right]\!\!, \label{eqn:regret_Mbnd}\end{align} where $T_{i,j}(n) $ is
the number  of slots where user $j$ selects channel $i$ for sensing, $M(n)$ is the  number of  collisions faced by  the users in  the $U$-best channels  in $n$ slots, $\Delta(i,j)=\mu(i)-\mu(j)$ and $\mu(1^*)$ is the highest mean availability.
\end{proposition}

\bprf  See Appendix~\ref{proof:bounds}.\eprf

In the subsequent sections, we propose distributed learning and access policies and   provide regret guarantees for the policies  using the upper bound in (\ref{eqn:regret_Mbnd}). The lower bound in \eqref{eqn:regret_Mlowerbnd} can be used to   derive lower bound on regret for any uniformly-good policy. 

The first term   in  (\ref{eqn:regret_Mbnd}) represents the lost transmission opportunities due to selection of   $U$-worst channels (with  lower mean availabilities), while the second term represents performance loss due to collisions among the users in the $U$-best channels. The first term in \eqref{eqn:regret_Mbnd}  decouples among the different users and can be analyzed solely through the marginal distributions of the $g$-statistics at the users. This in turn,  can be  analyzed by manipulating the classical results on multi-armed bandits \cite{Agrawal:95AAP,Auer&etal:02ML}.    On the other hand, the second term in \eqref{eqn:regret_Mbnd}, involving collisions in the $U$-best channels, requires the joint distribution of the $g$-statistics at different users which are correlated variables. This is intractable to analyze directly and we develop techniques to bound this term.
 
\subsection{$\randalloc: $ Distributed  Learning and Access}

We present the $\randalloc$ policy in Algorithm~\ref{algo:randalloc}.  
Before describing this policy, we make some simple observations. If each user implemented the single-user policy in Algorithm~\ref{algo:singleuser}, then it would result in collisions, since all the users target the best channel. When there are multiple users and there is no direct communication among them, the users need to randomize   channel access in order to avoid collisions. At the same time, accessing the $U$-worst channels needs to be avoided since they contribute to regret. Hence, users can avoid collisions by  randomizing access over the $U$-best channels, based on their estimates of the channel ranks.  However, if the users randomize in every slot, there is a finite probability of collisions in every slot and this results in a linear growth of regret with the number of time slots.  Hence,  the users need to converge to a collision-free configuration to ensure that the regret is logarithmic.   

In Algorithm~\ref{algo:randalloc},  there is adaptive randomization based on feedback regarding  the previous transmission. Each user randomizes {\em only} if there is a collision in the previous slot; otherwise,   the previously generated random rank for the user is retained. The estimation for the channel ranks is through the $g$-statistic, on  lines similar to the single-user case.


\begin{algorithm}
{\bf Input:} $\{\bar{X}_{i,j}(n)\}_{i=1,\ldots,C}:$ Sample-mean availabilities at user $j$  
 after $n$ rounds, $g_j(i;n)$: statistic based on
$\bar{X}_{i,j}(n)$,  
$\sigma(T;\statbf_j(n))$: index of $T^{\tha}$ highest entry in 
$\statbf_j(n)$.\\  $\iscollide_j(i;n)$: indicator of collision at $n^{\tha}$ slot at channel $i$
\\ Init:  Sense  in each channel once, $n
\leftarrow C$, $\currank\leftarrow 1$,  $\iscollide_j(i;m)\!\leftarrow\! 0$
\\ Loop: $n\leftarrow n+1$ 
\\{\bf if} $\iscollide_j(\curchannel;n-1)= 1$ {\bf then}\\
Draw a new $\currank\sim\unif(U)$ \\ {\bf end if}
\\ Select channel for sensing. If free, transmit.  \\$\curchannel\leftarrow \sigma(\currank;\statbf_j(n))$. 
\\ {\bf If}  collision $\iscollide_j(\curchannel;m)\leftarrow 1$, {\bf Else} $0$.
\caption{Policy $\randalloc(U,C,\statbf_j(n))$ for each user $j$ 
under $U$  users,  $C$ channels and statistic $\statbf_j(n)$.} 
\label{algo:randalloc}
\end{algorithm}

\subsection{Regret Bounds under $\randalloc$}

It is easy to see that the $\randalloc$ policy  ensures that the users are allocated orthogonally to the $U$-best channels as the number of transmission slots goes to infinity. The regret bounds on $\randalloc$ are however not immediately clear and we provide guarantees  below.
  
We first provide a logarithmic  upper bound\footnote{Note that the
bound on $\Ebb[T_{i,j}(n)]$ in \eqref{eqn:tij_randalloc} holds for
user $j$ even if   the other users  are using a policy other than
$\randalloc$. But on the other hand, to analyze   the  number of
collisions $\Ebb[M(n)]$ in \eqref{eqn:M}, we need  every user to
implement $\randalloc$.} on the number of slots spent by each user
in any $U$-worst channel. Hence, the first term in the bound on
regret in \eqref{eqn:regret_Mbnd} is also logarithmic.

\bl[Time Spent in $U$-worst Channels]\label{lemma:uworstrandalloc}
Under the $\randalloc$ scheme in Algorithm~\ref{algo:randalloc}, the total
time spent by any user $j=1,\ldots, U$, in any $i\in U$-worst
channel is given by \beq \Ebb[T_{i,j}(n)] \leq
\sum_{k=1}^U\Bigl[\frac{ 8 \log n}{\Delta(i,k^*)^2} +
1+\frac{\pi^2}{3}\Big].\label{eqn:tij_randalloc}\eeq 
\el

\bprf The proof is on lines  similar to the proof for
Theorem~\ref{thm:lowerbnd_cent}, given in
Appendix~\ref{proof:lowerbnd_cent}. \eprf


We now focus on analyzing the number of collisions $M(n)$ in the
$U$-best channels. We first give a result on the expected number of
collisions in the ideal scenario where each user has  perfect
knowledge of the channel availability statistics $\mubf$. In this case, the
users attempt to  reach an orthogonal (collision-free) configuration by uniformly randomizing over the $U$-best channels. 

The stochastic process  in this case is a finite-state Markov chain. A state in this Markov chain corresponds to a configuration of  $U$ number of (identical) users in $U$ number of channels.  The number of states in the Markov chain is the number of {\em compositions} of $U$, given by $ \binom{2U-1}{U}$  \cite[Thm. 5.1]{Bona:book}. The orthogonal configuration corresponds to the absorbing state. For any other state, consisting of more than one user or no user in any of the channels, the transition probability to any state of the Markov chain (including self transition probability) is uniform. For a state, where certain channels have exactly one user, there are only transitions to states which consist of at least one user in that channel and the transition probabilities are uniform.   Let $\Upsilon(U,U)$ denote the maximum time to absorption in the above Markov chain starting from any initial distribution. We have the following result
  
\bl[\# of Collisions Under Perfect Knowledge]\label{lemma:Pi}The
expected number of collisions under  $\randalloc$ scheme in
Algorithm~\ref{algo:randalloc}, assuming that each user has  perfect
knowledge of the mean channel availabilities $\mubf$, is given by
\begin{align}\nn  \Ebb[M(n);\randalloc(U,C,\mubf)]&\leq U\Ebb[\Upsilon(U,U)]\\&   \leq  U\left[
\binom{2U-1}{U}\!-\!1\right].\label{eqn:Pi}\end{align}\el

\bprf See Appendix~\ref{proof:Pi}.\eprf

The above result states that there is at most a finite number of expected collisions, bounded by $U\Ebb[\Upsilon(U,U)]$ under perfect
knowledge of $\mubf$. In contrast, recall from the previous section,
that there are no collisions under perfect knowledge of $\mubf$ in
the presence of pre-allocated ranks. Hence, $U\Ebb[\Upsilon(U,U)]$ represents a bound on the additional regret due to the lack of direct communication among the users to negotiate their ranks.

We  use the result of Lemma \ref{lemma:Pi} for analyzing the number
of collisions under distributed learning of the unknown
availabilities $\mubf$ as follows: if we show that  the users are
able to learn the correct order of the different channels with only
logarithmic regret then   only an additional finite expected number
of collisions occur before reaching an  orthogonal configuration.

Define $T'(n;\randalloc)$ as the number of slots where any one of the top-$U$ estimated ranks of the channels at some user  is wrong under $\randalloc$ policy. 
Below we prove that its expected value is logarithmic in the number of transmissions.

\bl[Wrong Order of
$g$-statistics]\label{lemma:numintfrandalloc}Under the $\randalloc$
scheme in Algorithm~\ref{algo:randalloc},   \beq \Ebb[T'(n;\randalloc)] \leq
U\sum_{a=1}^U \sum_{b=a+1}^C  \left[ \frac{8 \log n}{\Delta(a^*,b^*)^2}+\!1\! + \frac{\pi^2}{3}\right]\!
.\!\!\!\!\!\label{eqn:tdashj}\eeq   \el

\bprf  See Appendix~\ref{proof:numintfrandalloc}. \eprf

We now provide an upper bound on the number of collisions $M(n)$ in
the $U$-best channels by incorporating the above result on
$\Ebb[T'(n)]$, the result on  the average number of slots
$\Ebb[T_{i,j}]$ spent in the $U$-worst channels in Lemma
\ref{lemma:uworstrandalloc} and  the average number of collisions
$U\Ebb[\Upsilon(U,U)]$ under perfect knowledge of $\mubf$ in  Lemma
\ref{lemma:Pi}.

\bt[Logarithmic Number of Collisions Under
$\randalloc$]\label{thm:intfrandalloc} The expected number of
collisions in the $U$-best channels under $\randalloc(U,C,
\bfg^{\mbox{\tiny MEAN}} )$ scheme satisfies \beq \Ebb[M(n)] \leq 
U(\Ebb[\Upsilon(U,U)]+1)\,\Ebb[T'_j(n)].\label{eqn:M}\eeq 
Hence, from \eqref{eqn:tij_randalloc}, \eqref{eqn:tdashj} and
\eqref{eqn:Pi}, $M(n) = O(\log n)$.\et

\bprf See Appendix~\ref{proof:thmintfrandalloc}. \eprf


Hence, there are only logarithmic number of expected collisions
before the users settle in the orthogonal channels. Combining this
result with Lemma \ref{lemma:uworstrandalloc} that the number of
slots spent in the $U$-worst channels is also logarithmic, we
immediately have one of the main results of this paper that the sum
regret under distributed learning and access is logarithmic.

\bt[Logarithmic Regret Under $\randalloc$]\label{thm:randalloc}The
policy $\randalloc(U,C,\bfg^{\mbox{\tiny MEAN}} )$ in
Algorithm~\ref{algo:randalloc} has $\Theta(\log n)$   regret.\et

\bprf Substituting  \eqref{eqn:M} and \eqref{eqn:tij_randalloc} in
\eqref{eqn:regret_Mbnd}.\eprf

Hence, we prove that distributed learning and channel access  among multiple
secondary users is possible with logarithmic regret without any
explicit communication among the users. This implies that the number
of lost opportunities for successful transmissions at all secondary users is only
logarithmic in the number of transmissions, which is negligible when
there are large number of transmissions.

We have so far focused on designing schemes that maximize system or
social throughput. We now briefly discuss the fairness for an
individual user under $\randalloc$. Since $\randalloc$ does not
distinguish any of the users,  in the sense that each user
has equal probability of ``settling" down in one of the $U$-best
channels while  experiencing only logarithmic regret in doing so.
Simulations in Section~\ref{sec:num} (in Fig.\ref{fig:fairness})
demonstrate this phenomenon.
%


\section{Distributed Learning and Access under Unknown Number of
Users}\label{sec:unknown}

We have so far assumed that the number of secondary users is known, and is required for the implementation of  the   $\randalloc$ policy. In practice, this entails initial   announcement from each of the secondary users to indicate their presence  in the cognitive network. However, in a truly distributed setting without any 
information exchange among the users, such an announcement may not be possible. 

In this section, we consider the scenario, where the number of users $U$ is unknown (but fixed throughout the duration of transmissions and $U\leq C$, the number of channels). In this case, the policy needs to estimate the number of secondary users in the system, in addition to learning the channel availability statistics and designing channel access rules based on collision feedback. Note that if the policy assumed the worst-case scenario that $U=C$, then the regret grows linearly since $U$-worst channels are selected a large number of times for sensing.

\subsection{Description of $\unknownalloc$ Policy}

We now propose a policy $\unknownalloc$ in Algorithm~\ref{algo:unknownalloc}. This policy incorporates two functions in each transmission slot, viz.,  execution of the $\randalloc$ policy in Algorithm~\ref{algo:randalloc}, based on the current estimate of the number of users $\widehat{U}$, and  updating  of the estimate $\widehat{U}$ based on  the number of collisions experienced by the user.
  
\begin{algorithm}\begin{enumerate}\item{\bf Input:} $\{\bar{X}_{i,j}(n)\}_{i=1,\ldots,C}:$ Sample-mean availabilities at user $j$,  $g_j(i;n)$: statistic based on
$\bar{X}_{i,j}(n)$, \\ 
$\sigma(T;\statbf_j(n))$: index of $T^{\tha}$ highest entry in 
$\statbf_j(n)$.\\ $\iscollide_j(i;n)$: indicator of collision at $n^{\tha}$ slot at channel $i$\\ $\widehat{U}$: current estimate of the number of 
users.\\  $n$: horizon (total number of slots for transmission) 
\item Init:  Sense   each channel once, $m \leftarrow C$, 
$\currank\leftarrow 1$, $\widehat{U}\!\leftarrow \!1$, 
 $\iscollide_j(i;m)\!\leftarrow\! 0$ for all $i=1,\ldots, C$   \item 
Loop: $m\leftarrow m+1$, stop when $m =n$. 
\item \label{algoline:rand_unknownalloc}  {\bf If} $\iscollide_j(\curchannel;m-1)= 1$ {\bf then}\\ Draw a new $\currank\sim\unif(\widehat{U})$.   {\bf end if}
\\ Select channel for sensing.  If free, transmit. \\ $\curchannel\leftarrow \sigma(\currank;\statbf_j(m))$
\item $\iscollide_j(\curchannel;m)\leftarrow 1$  if collision, $0$ o.w. \item\label{algoline:increment_unknownalloc} {\bf If} $\sum_{a=1}^m\sum_{k=1}^{\widehat{U}}\iscollide_j(\sigma(k;\statbf_j(m));a) > \xi(n;\widehat{U}))$ {\bf then}\\
$\widehat{U}  \leftarrow \widehat{U}  +1,$ $\iscollide_j(i;a)\leftarrow 0$, $i=1,\ldots C$, $a=1,\ldots, m$. {\bf end if}
\een\caption{Policy $\unknownalloc(n,C,\statbf_j(m), \xi)$
for each user $j$ under $n$ transmission slots (horizon length),   $C$ channels, 
statistic $\statbf_j(m)$ and threshold functions $\xi$.} 
\label{algo:unknownalloc}\end{algorithm}

The updating is  based on the  idea that if there is  
under-estimation of $U$ at all the users ($\widehat{U}_j< U$ at all the users $j$), collisions necessarily build up and  the collision count serves as a criterion for incrementing   $\widehat{U}$.   This is because after a long learning period,  the users    learn the true ranks of the channels, and target the same set of channels. However, when there is under-estimation, the number of users exceeds the number of channels targeted by the users. Hence, collisions among the users accumulate, and can be used as a test for incrementing  $\widehat{U}$.
 
Denote the collision count used by $\unknownalloc$ policy as\beq\label{eqn:Phi} \Phi_{k, j}(m) := 
\sum_{a=1}^m\sum_{b=1}^{k}\iscollide_j(\sigma(b;\statbf_j(m));a). \eeq which is the total number of collisions experienced by user $j$ so far (till the $m^{\tha}$ transmission slot) in the  top-$\widehat{U}_j$ channels, where the ranks of the channels are estimated using the $g$-statistics. The collision count is tested against a  threshold  $\xi(n;\widehat{U}_j)$, which is a function of the horizon length\footnote{In this section, we assume that the users are aware of the horizon length $n$ for transmission. Note that this is not a limitation and can be extended to case of unknown horizon length as follows: implement the algorithm by fixing horizon lengths to $n_0, 2n_0 , 4n_0\ldots$ for a fixed $n_0 \in \Nbb$ and discarding estimates from previous stages.} and current estimate $\widehat{U}_j$. When the threshold is exceeded, $\widehat{U}_j$ is incremented, and the collision samples collected so far are discarded (by setting them to zero)  (line~\ref{algoline:increment_unknownalloc} in 
Algorithm~\ref{algo:unknownalloc}).
 
\subsection{Regret Bounds under $\unknownalloc$}

We analyze regret bounds under the $\unknownalloc$ policy, where the regret is defined in \eqref{eqn:regret_collision}.  Let the maximum threshold function for the number of consecutive collisions under $\unknownalloc$ policy be denoted by
  \beq \label{eqn:xi_star} \xi^*(n;U) \defeq
\max_{k=1,\ldots, U}\xi(n;k).\eeq We   prove that the  
$\unknownalloc$ policy has $O(\xi^*(n;U))$ regret when $\xi^*(n;U) = \omega(\log n)$, and where $n$ is   the number of transmission slots.

The proof for the regret bound under $\unknownalloc$ policy   consists of two main parts: we    prove bounds on  regret conditioned on the  event that none of the users over-estimate $U$. Second, we show that the probability of over-estimation at any of the users goes to zero   asymptotically. Combined together, we obtain the regret bound for  $\unknownalloc$ policy. 

Note that  in order to have small regret, it is crucial that none of the users over-estimate $U$. This is because when there is over-estimation, there is a  finite probability of selecting the $U$-worst channels    even upon learning the true ranks of the channels. Note that regret is incurred whenever a U-worst channel is selected since under perfect knowledge this channel would not be selected. Hence, under over-estimation, the regret grows linearly in the number of transmissions.

In a nutshell, under the $\unknownalloc$ policy, the decision to increment  the estimate $\widehat{U}$ reduces to a hypothesis-testing problem with  hypotheses $\Hc_0$: number of users is less than or equal to the current estimate and $\Hc_1$: number of users is greater than the current estimate. In order to have a sub-linear regret, the false-alarm probability (deciding $\Hc_1$ under $\Hc_0$) needs to decay asymptotically. This is ensured by selecting appropriate thresholds $\xi(n)$ to test against the collision counts obtained through feedback.

\subsubsection*{Conditional Regret}

We now give the result for the first part.  Define the ``good event'' $\Cc(n;U)$ that 
none of the users over-estimates $U$ under 
$\unknownalloc$ as\beq\label{eqn:good}\Cc(n;U)\defeq\{\bigcap_{j=1}^U\widehat{U}^\est_j(n) \leq U\}.\eeq 


The regret conditioned on $\Cc(n;U)$, denoted by $R(n;\mubf, 
U, \unknownalloc)| \Cc(n;U) $,  is given by
\[  n \sum_{k=1}^U \mu(k^*)  -\sum_{i=1}^C \sum_{j=1}^U \mu(i)
\Ebb[V_{i,j}(n)|\Cc(n;U)] ,\]  where $V_{i,j}(n)$ is the number of times  
that   user $j$ is the sole user of channel $i$. 
Similarly, we have conditional expectations of $\Ebb[T_{i,j}(n)|\Cc(n;U)]$ and of the number of collisions in $U$-best channels, given by $\Ebb[M(n)|\Cc(n;U)]$. We now show that the regret 
conditioned on $\Cc(n;U)$ is $O(\max(\xi^*(n;U), \log n) )$.

\bl\label{lemma:cond_regret}\emph{(Conditional Regret): }When all
the $U$ secondary users implement $\unknownalloc$ policy, we have
for all $ i \in U$-worst channel and each user $j=1,\ldots, U$,\beq
\label{eqn:cond_uworst} \Ebb[T_{i,j}(n)|\Cc(n)]\leq
\sum_{k=1}^U\left[\frac{8\log
n}{\Delta(i,k^*)^2}+1+\frac{\pi^2}{3}\right].\eeq The conditional
expectation on number of collisions $M(n)$ in the $U$-best channel
satisfies \beq\label{eqn:cond_m} \Ebb[M(n)|\Cc(n;U)]\leq U \sum_{k=1}^U \xi(n;k) \leq U^2 \xi^*(n;U).
\eeq From \eqref{eqn:regret_Mbnd}, we have $R(n)|\Cc(n;U)$ is  $ O( \max(\xi^*(n;U), \log n))$ for 
any $n\in \Nbb$.\el

\bprf   See Appendix~\ref{proof:cond_regret}.\eprf
 
\subsubsection*{Probability of Over-estimation}

We now prove that none of the users over-estimates\footnote{Note that $\unknownalloc$ policy automatically ensures that all the users do not under-estimate $U$, since it increments $\widehat{U}$ based on collision estimate. This implies that the probability of the event that all the users under-estimate $U$ goes to zero asymptotically.}  $U$ under $\unknownalloc$ policy,
\ie the probability of the event $\Cc(n;U) $ in \eqref{eqn:good} approaches one as $n \to \infty$, when the thresholds $\xi(n;\widehat{U})$ for testing against the collision count  are chosen appropriately (see line~\ref{algoline:increment_unknownalloc} in 
Algorithm~\ref{algo:unknownalloc}). Trivially, we can set $\xi(n;1)=1$ 
since a single collision is enough to indicate that there is more than one user. For any other $k>1$, we choose    functions $\xi$  satisfying \beq \label{eqn:cond_xi} 
\xi(n;k) = \omega(\log n), \quad \forall k>1.\eeq We prove that the above condition ensures that over-estimation does not occur. 

Recall that $T'(n;\unknownalloc)$ is the number of slots where any one of the top-$U$ estimated ranks of the channels at some user  is wrong under $\unknownalloc$ policy.  We  show that $\Ebb[T'(n)]$ is $O(\log n)$.

\bl[Time spent with wrong estimates]\label{lemma:Tdashdash} The expected number of slots  where any of the top-$U$ estimated ranks of the channels at any user  is wrong under $\unknownalloc$ policy satisfies \beq \label{eqn:Tdashdash} \Ebb[T'(n)]\leq U\sum_{a=1}^U \sum_{b=a+1}^C  \left[ \frac{8 \log n}{\Delta(a^*,b^*)^2}+\!1\! + \frac{\pi^2}{3}\right]. \eeq\el

\bprf The proof is on the lines of Lemma~\ref{lemma:numintfrandalloc}\eprf

Recall the definition of $\Upsilon(U,U)$ in the previous section, as the maximum time to absorption   starting from any initial distribution of the finite-state Markov chain, where the states correspond to different user configurations and the absorbing state corresponds to the collision-free configuration. We now generalize the definition to $\Upsilon(U,k)$, as the time to absorption in a new Markov chain, where the state space   is the set of configurations of $U$  users in $k$  channels, and the transition probabilities are defined on similar lines. Note that $\Upsilon(U,k)$ is almost-surely finite when $k\geq U$  and $\infty$ otherwise (since there is no absorbing  state in the latter case). 

We now bound the maximum value of the collision count $\Phi_{k,j}(m)$ under $\unknownalloc$ policy in \eqref{eqn:Phi} using $T'(m)$, the total time spent with wrong channel estimates,  and $\Upsilon(U,k)$, the time to absorption in the Markov chain. Let $\leqst$ denote the stochastic order for two random variables \cite{Shaked&Shanthikumar:book}.

\begin{proposition}\label{prop:numcoll}The maximum collision count  in \eqref{eqn:Phi} over all users under the $\unknownalloc$ policy satisfies\beq\label{eqn:numcoll} \max_{j =1,\ldots, U}
 \Phi_{k,j}(m)\leqst (T'(m)+1) \Upsilon(U,k),\quad \forall m \in \Nbb.\eeq  \end{proposition}

\bprf The proof is on the lines of Theorem~\ref{thm:intfrandalloc}. See Appendix~\ref{proof:numcoll}. \eprf

We now prove that the probability of over-estimation goes to zero asymptotically.

\bl[No Over-estimation  Under
$\unknownalloc$]\label{lemma:underestimate} For threshold functions
satisfying \eqref{eqn:cond_xi}, the event $\Cc(n;U) $ in
\eqref{eqn:good} satisfies \beq\lim_{n \to
\infty}\Pbb[\Cc(n;U)]=1,\label{eqn:underestimate}\eeq and hence, none
of the  users over-estimates $U$ under 
$\unknownalloc$ policy.  \el

\bprf  See Appendix~\ref{proof:underestimate}.\eprf

We now give the main result of this section that $\unknownalloc$ has
slightly more than logarithmic regret asymptotically and this
depends on the threshold function $\xi^*(n;U)$ in
\eqref{eqn:xi_star}.

\bt[Asymptotic  Regret Under $\unknownalloc$]With threshold 
functions $\xi$ satisfying conditions in \eqref{eqn:cond_xi},   the 
policy $\unknownalloc(n,C,\statbf_j(m), \xi)$ in 
Algorithm~\ref{algo:unknownalloc} satisfies \beq 
\label{eqn:regret_bound_unknown}\limsup_{n \to \infty} 
\frac{R(n;\mubf, U, \unknownalloc)}{\xi^*(n;U)} <\infty. \eeq
 \et

\bprf From Lemma~\ref{lemma:cond_regret} and 
Lemma~\ref{lemma:underestimate}.\eprf

Hence, the regret  under the proposed $\unknownalloc$ policy  is  
$O(\xi^*(n;U))$ under fully  decentralized setting without the 
knowledge of number of users when $\xi^*(n;U)= \omega(\log n)$.  Hence,  $O(f(n)\log  
n)$ regret is achievable  for all functions $f(n) \to \infty $ as  $n \to \infty$. The question of whether logarithmic regret is possible under unknown number of users is of interest. 


Note the difference between $\unknownalloc$ policy in 
Algorithm~\ref{algo:unknownalloc} under unknown number of users with 
$\randalloc$ policy with known number of users in Algorithm~\ref{algo:randalloc}. 
The  regret under $\unknownalloc$ is $O(f(n)\log n )$ for any function $f(n) = \omega(1)$, while 
it is $O(\log n)$ under $\randalloc$ policy. Hence, we are able to quantify the degradation of performance when the number of users is unknown.

\section{Lower Bound \& Effect of Number of Users}\label{sec:lowerbound}
\subsection{Lower Bound For Distributed Learning \& access}

We have so far designed distributed learning and access policies with provable bounds on regret. We now discuss the relative performance of these policies, compared to the optimal learning and access policies. This is accomplished by noting a lower bound on regret for any {\em uniformly-good} policy, first derived in \cite{Liu&Zhao:09Arxiv} for a   general class 
of uniformly-good time-division policies. We restate the result below.

%
%

\bt[Lower Bound \cite{Liu&Zhao:09Arxiv}]\label{thm:lowerbnd}For any uniformly good distributed learning and access policy $\policy$,   the sum regret in \eqref{eqn:regret_def} satisfies\beq \liminf_{n \to \infty}\frac{R(n;\mubf,U,\policy)}{\log n} \geq
  \sum_{i \in U \mbox{\scriptsize-worst}} \sum_{j=1}^U \frac{\Delta(U^*,i)
}{D(\mu_i,\mu_{j^*})} .\label{eqn:lowerbndregret}\eeq  \et

The lower bound derived in \cite{Anantharam&etal:87TranAC} for
centralized learning and access  holds for distributed
learning and access considered here. But a better lower bound
is obtained above by considering the distributed nature of learning.
The lower bound for distributed policies   is  worse than the bound for the centralized policies in \eqref{eqn:lowerbndregret_central}. This is because each user independently learns the channel availabilities $\mubf$ in a distributed policy, whereas sensing  decisions from all the users are used for learning in a centralized policy.   

Our distributed learning and access policy $\randalloc$ matches the lower bound on regret in \eqref{eqn:regret_Mbnd} in the order $(\log n)$ 
but the scaling factors are different. It is not clear if the regret 
lower bound in \eqref{eqn:lowerbndregret} can be achieved  by any 
policy under no explicit information exchange and is a topic for 
future investigation.

\subsection{Behavior with Number of Users}\label{sec:extensions}

We have so far analyzed the sum regret under our policies  under a fixed number of users $U$. We now analyze the behavior of regret growth as $U$ increases while keeping the number of channels $C>U$ fixed.

\bt[Varying Number of Users]\label{thm:varyinguc} When the number of
channels $C$ is fixed and the number of users $U<C$ is varied, the
sum regret under centralized learning and access $\cent$ in
\eqref{eqn:regret_cent_opt} decreases as $U$ increases while the
upper bounds on the sum regret under    $\randalloc$
in \eqref{eqn:regret_Mbnd} monotonically increases with $U$. \et

\bprf The proof involves analysis of \eqref{eqn:regret_cent_opt}
   and  \eqref{eqn:regret_Mbnd}. To prove
that the sum regret under centralized learning and access in
\eqref{eqn:regret_cent_opt} decreases with the number of users $U$,
it suffices to show that for $i \in U$-worst channel,\[
\frac{\Delta(U^*, i)}{D(\mu_i,\mu_{U^*})}\] decreases as $U$
increases. Note that $\mu(U^*)$ and $ D(\mu_i,\mu_{U^*})$ decrease
as $U$ increases. Hence, it suffices to show that
\[\frac{\mu(U^*)}{D(\mu_i,\mu_{U^*})}\] decreases with $U$. This is true since
its derivative with respect to $U$ is negative.

For the upper bound on regret under $\randalloc$ in
\eqref{eqn:regret_Mbnd}, when $U$ is increased, the number of
$U$-worst channels decreases and hence, the first term in
\eqref{eqn:regret_Mbnd} decreases. However, the second term
consisting of collisions $M(n)$ increases to a far greater
extent.\eprf

Note that the above results is for the upper bound  on regret  under the $\randalloc$ policy and not the regret itself.  Simulations in Section~\ref{sec:num} reveal that the actual regret also increases
with $U$. Under the centralized scheme $\cent$, as $U$ increases, the  number
of $U$-worst channels decreases. Hence, the regret decreases, since
there are less number of possibilities of making bad decisions.
However, for distributed schemes although this effect exists, it is
far outweighed by the increase in regret due to the increase in
collisions among the $U$ users.

In contrast, the distributed  lower bound in
\eqref{eqn:lowerbndregret} displays anomalous behavior with $U$
since it fails to account for collisions among the users. Here, as
$U$ increases there are two competing effects: a decrease in  regret
due to decrease in the number of $U$-worst channels and an increase
in regret due to increase in the number of users visiting these
$U$-worst channels.

\section{Numerical Results}\label{sec:num}

 \begin{figure*}[t]\subfloat[a][Normalized regret  $\tfrac{ R(n)}{\log
n}$ vs. $n$ slots.\\  $U= 4$ users,  $C= 9$
channels.]{\label{fig:algocomp}\begin{minipage}{2.3in}
\bc\bp\psfrag{No. of plays}[c]{}\psfrag{Regret}[c]{}\psfrag{Regret
Coefficient (Rand)}[l]{\tiny Random Allocation Scheme}\psfrag{Regret
Coefficient (Cent)}[l]{\tiny Central Allocation
Scheme}\psfrag{Regret Coefficient (Pre)}[l]{\tiny Pre-Allocation
Scheme}\psfrag{Theoretical Lower Bound (Distributed)}[l]{\tiny
Distributed Lower Bound}\psfrag{Theoretical Lower Bound
(Centralized)}[l]{\tiny
 Centralized Lower Bound}
 \includegraphics[width=2.3in,height=1.5in]{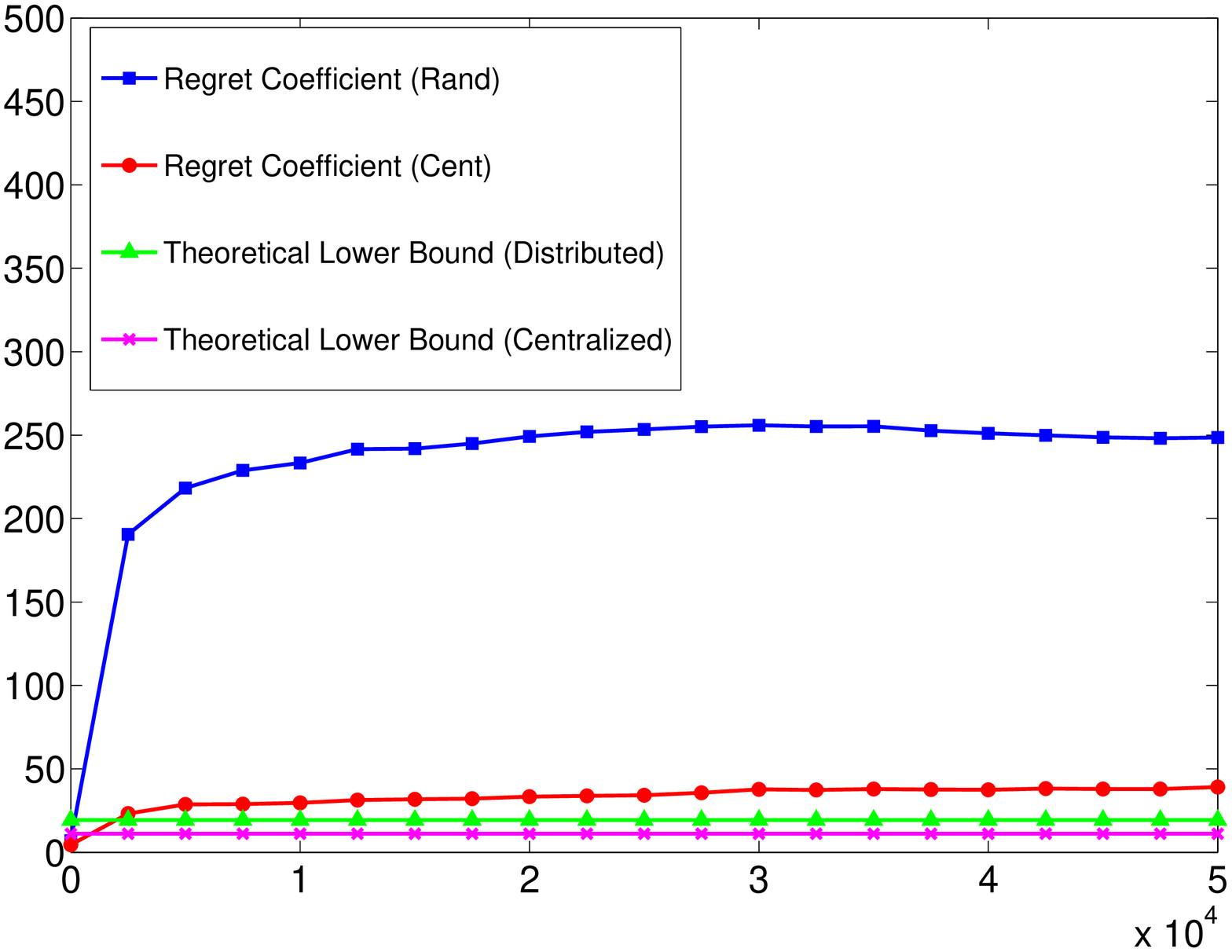}
\ep\ec
\end{minipage}}\hfil\subfloat[b][Normalized regret  $\tfrac{ R(n)}{\log
n}$   vs. $n$ slots.\\ $U=4$ users,  $C= 9$ 
channels.]{\label{fig:statcomp}\begin{minipage}{2.3in} 
\bc\bp\psfrag{No. of 
plays}[c]{}\psfrag{Regret}[c]{}\psfrag{Simulated 
Regret(g-mean)}[l]{\tiny $\randalloc$  with $g^{\mbox{\tiny MEAN}}$} 
\psfrag{Simulated Regret(g-opt)}[l]{\tiny $\randalloc$ with 
$g^{\mbox{\tiny OPT}}$}\psfrag{Theoretical Lower Bound on 
Regret}[l]{\tiny Distributed Lower Bound}
\includegraphics[width=2.3in,height=1.5in]{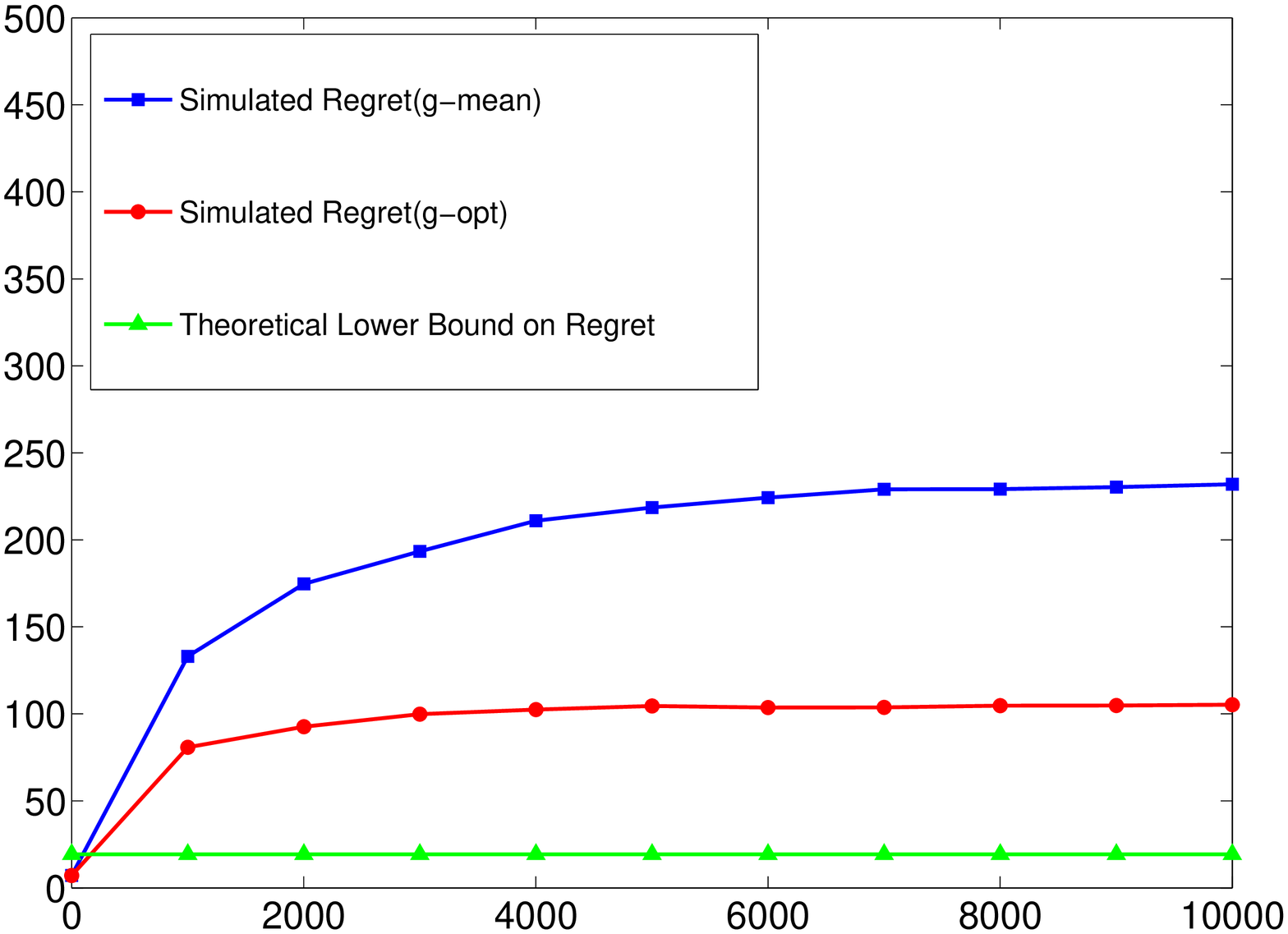}
\ep\ec
\end{minipage}} \hfil\subfloat[c][No. of collisions $M(n)$ vs. $n$ slots.\\  $U=4$ users, $C=9$ channels,
  $\randalloc$ policy.]{\label{fig:known}\begin{minipage}{2.3in}
\bc\bp\psfrag{No. of plays}[c]{}\psfrag{No. of
Collisions}[l]{}\psfrag{mu unknown}[l]{\tiny $\mubf$ unknown} 
\psfrag{No. of Collisions (Known)}[l]{\tiny $\mubf$ known: 
$U\Ebb[\Upsilon(U,U)]$}\psfrag{Upper bound on PiU}[l]{\tiny Upper Bound on 
$U\Ebb[\Upsilon(U,U)]$}
\includegraphics[width=2.3in,height=1.5in]{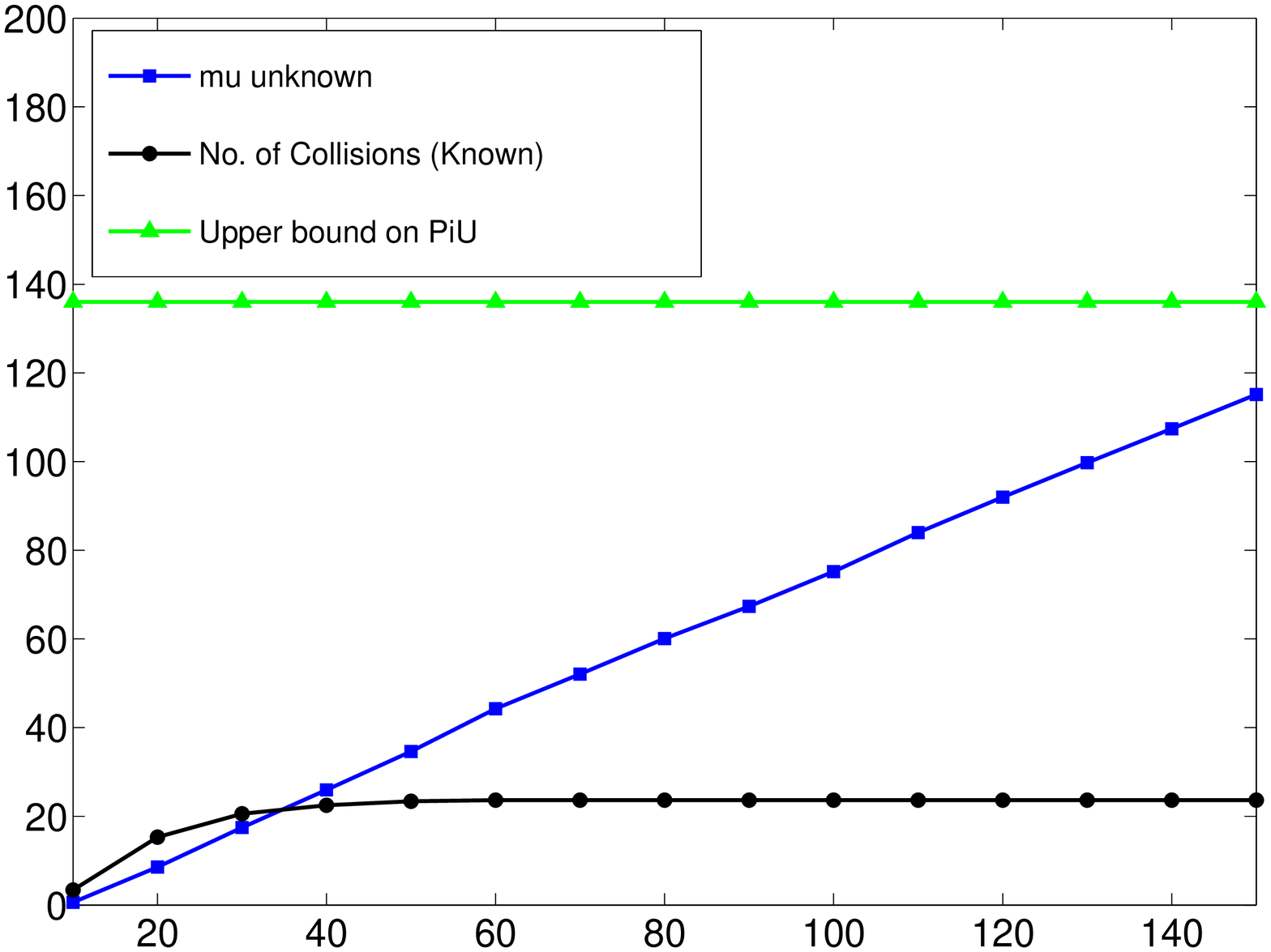}
\ep\ec
\end{minipage}}\caption{Simulation Results.
Probability of Availability $\mubf=[0.1,0.2,\ldots,
0.9]$.} \end{figure*}

 \begin{figure*}[t]
 \subfloat[a][Normalized regret  $\tfrac{ R(n)}{\log
n}$ vs.  $U$ users.\\  $C=9$ channels,  $n=2500$
slots.]{\label{fig:users}\begin{minipage}{2.3in} \bc\bp\psfrag{No. of plays}[c]{}
\psfrag{Regret}[c]{}\psfrag{Normalized Regret
(Rand)}[l]{\tiny Random Allocation Scheme}\psfrag{Normalized Regret
(Cent)}[l]{\tiny Central Allocation Scheme}\psfrag{Theoretical Lower
Bound (Distributed)}[l]{\tiny  Distributed Lower
Bound}\psfrag{Theoretical Lower Bound (Centralized)}[l]{\tiny
 Centralized Lower Bound}
\includegraphics[width=2.3in,height=1.5in]{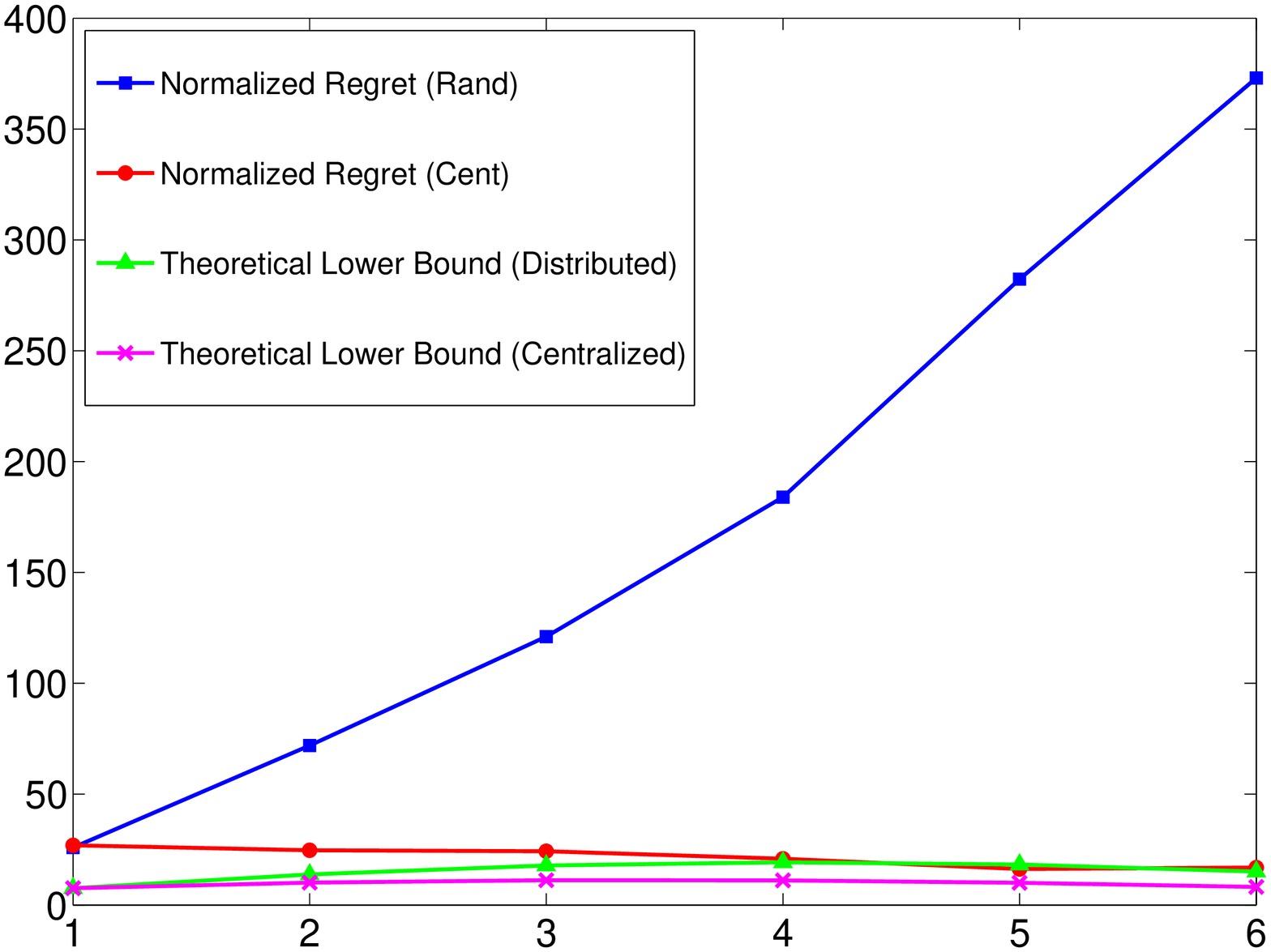}
\ep\ec
\end{minipage}}\hfil
 \subfloat[b][Normalized regret  $\tfrac{ R(n)}{\log
n}$ vs. $C$ channels.\\  $U=2$ users,  $n=2500$
slots.]{\label{fig:channels}\begin{minipage}{2.3in}
\bc\bp\psfrag{No. of
channels}[c]{}\psfrag{Regret}[c]{}\psfrag{Normalized Regret
(Rand)}[l]{\tiny Random Allocation Scheme}\psfrag{Normalized Regret
(Cent)}[l]{\tiny Central Allocation Scheme}\psfrag{Normalized Regret
(Pre)}[l]{\tiny Pre-Allocation Scheme}\psfrag{Theoretical Lower
Bound (Distributed)}[l]{\tiny  Distributed Lower
Bound}\psfrag{Theoretical Lower Bound (Centralized)}[l]{\tiny
 Centralized Lower Bound}\includegraphics[width=2.3in,height=1.5in]{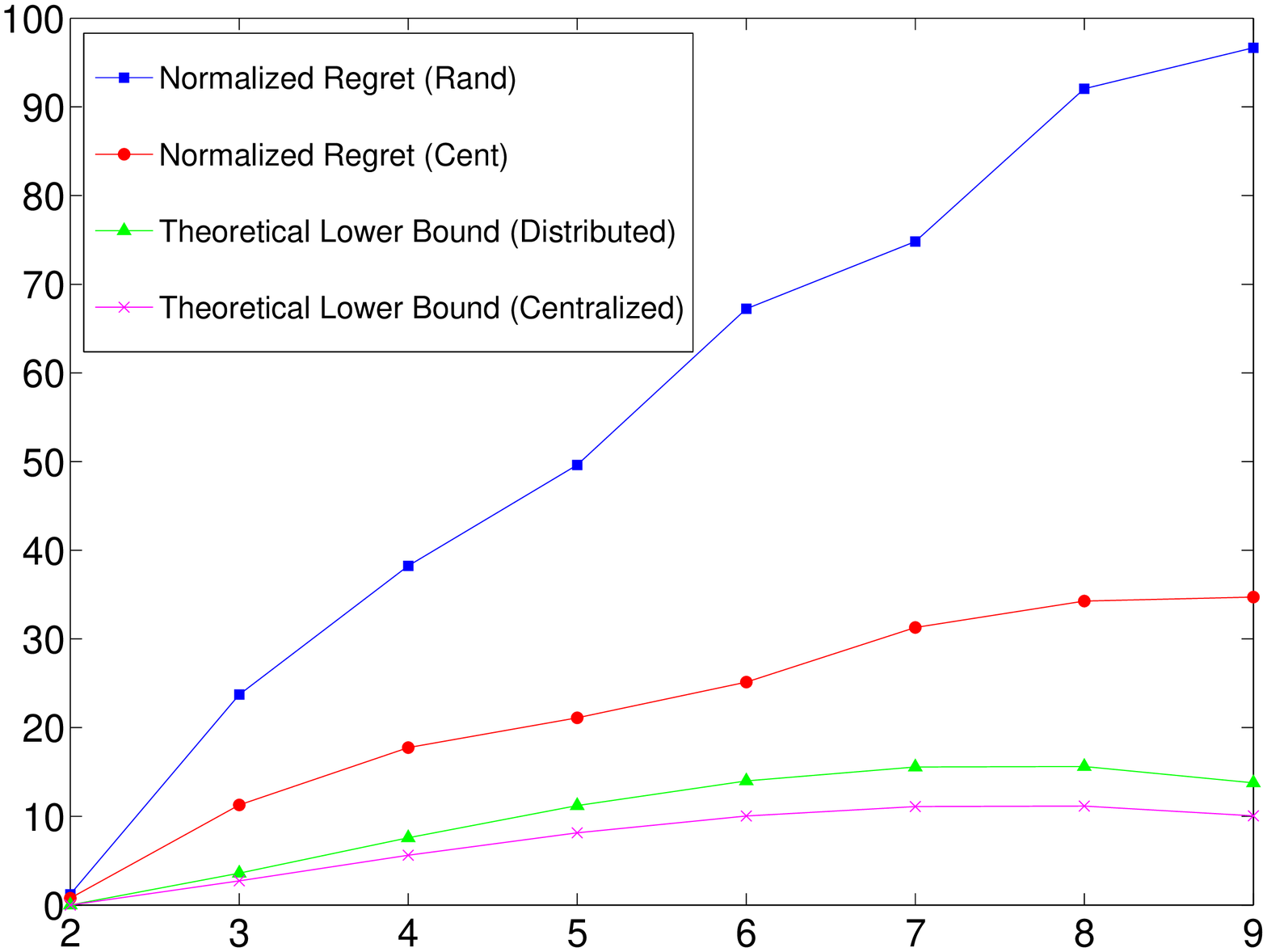}
\ep\ec
\end{minipage}}\subfloat[c][Normalized regret  $\tfrac{ R(n)}{\log
n}$ vs. $U$ users.\\ User-channel ratio $\frac{U}{C}=0.5$, $n=2500$
slots.]{\label{fig:upperbound}\begin{minipage}{2.3in}
\bc\bp\psfrag{No. of  users}[c]{} \psfrag{Norm.
Regret}[c]{}\psfrag{Normalized Regret (Rand)}[l]{\tiny Random
Allocation Scheme}\psfrag{Normalized Regret (Cent)}[l]{\tiny Central
Allocation Scheme}\psfrag{Normalized Regret (Pre)}[l]{\tiny
Pre-Allocation Scheme}\psfrag{Theoretical Lower Bound
(Distributed)}[l]{\tiny  Distributed Lower Bound}\psfrag{Theoretical
Lower Bound (Centralized)}[l]{\tiny
 Centralized Lower Bound}\includegraphics[width=2.3in,height=1.5in]{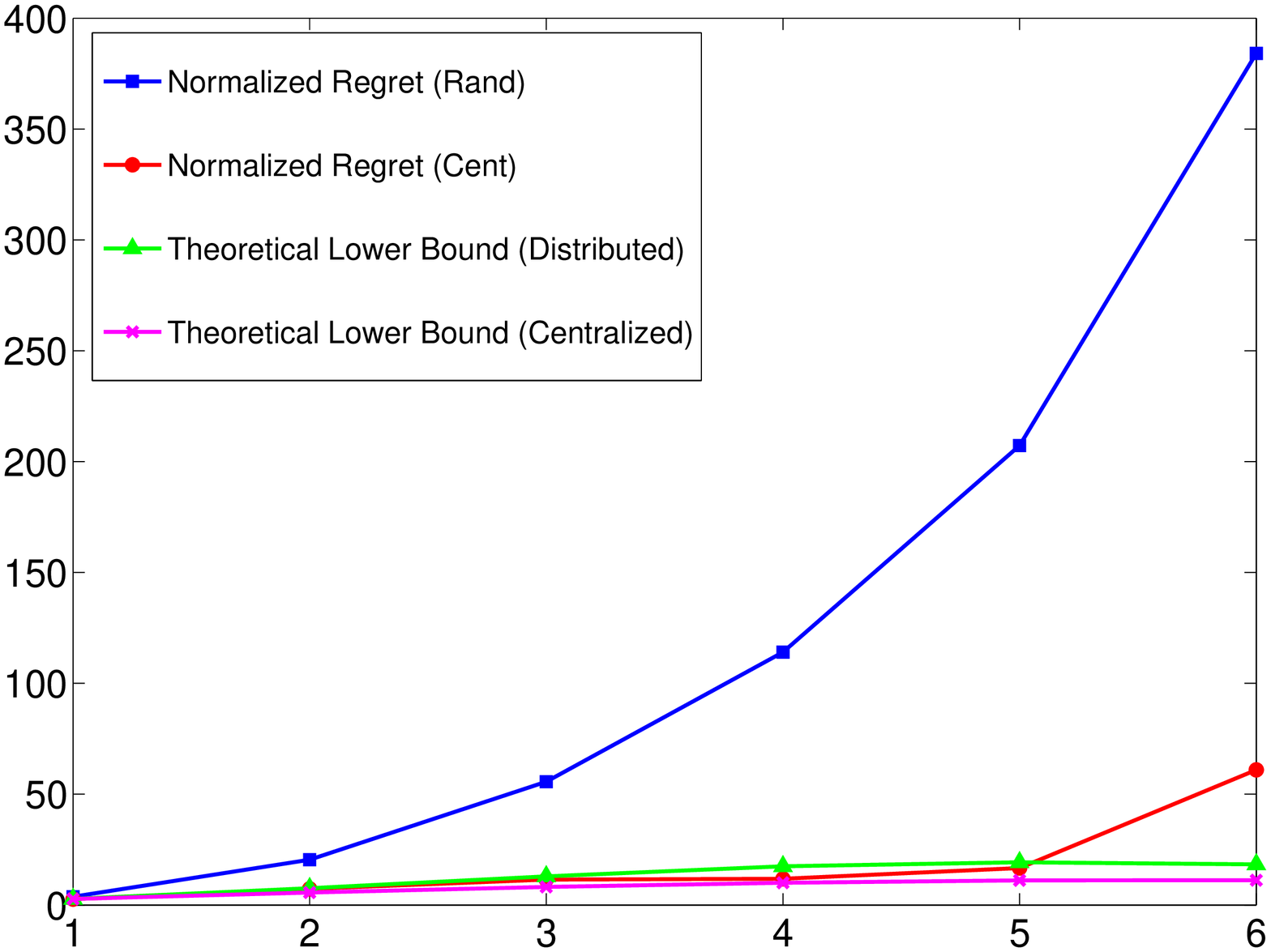}
\ep\ec
\end{minipage}}  \caption{Simulation Results.
Probability of Availability $\mubf=[0.1,0.2,\ldots,
0.9]$.} 
 \end{figure*}

We present simulations that vary the schemes and the number of users
and channels to verify the performance of the algorithms detailed
earlier.  We consider $C $= $9$ channels (or a subset of them when
the number of channels is varying) with probabilities of
availability characterized by Bernoulli distributions with evenly
spaced parameters ranging from $0.1$ to $0.9$. 
\vspace{0.05in}
\subsubsection*{Comparison of Different Schemes}
Fig.\ref{fig:algocomp} compares the regret under the centralized and
random allocation schemes in a scenario with $U$ = $4$ cognitive
users vying for access to the  $C=9$ channels. The theoretical
lower bound for the regret in the centralized case from Theorem
\ref{thm:lowerbnd_cent} and the distributed case from Theorem \ref{thm:lowerbnd}
are also plotted. The upper bounds on the random allocation scheme from
Theorem \ref{thm:randalloc} is not plotted here, since  the bounds
are loose especially as the number of users $U$ increases. Finding
tight upper bounds is a subject of future study.

As expected, centralized allocation has the least regret. Another 
important observation is the gap between the lower bounds on the 
regret and the actual regret in both the distributed and the 
centralized cases. In the centralized scenario, this is simply due 
to using the $g^{\mbox{\tiny MEAN}}$ statistic in 
\eqref{eqn:geomean} instead of the optimal $g^{\mbox{\tiny OPT}}$ 
statistic in \eqref{eqn:gopt}. However, in the distributed case, 
there is an additional gap since we do not account for collisions 
among the users. Hence, the schemes under consideration are  $O(\log 
n)$ and achieve order optimality although they are not optimal in 
the scaling constant. \vspace{0.05in}

\subsubsection*{Performance with Varying $U$ and $C$}
Fig.\ref{fig:users} explores the impact of increasing the number of
secondary users $U$ on the regret experienced by the different
policies while fixing the number of channels $C$.  With increasing
$U$, the regret   decreases for the centralized schemes and
increases for the distributed schemes, as predicted in
Theorem~\ref{thm:varyinguc}. The monotonic increase of regret under
random allocation $\randalloc$ is a result of the increase in the
collisions as $U$ increases. While the monotonic decreasing behavior
in the centralized case is because as the number of users increases,
the number of $U$-worst channels decreases resulting in lower
regret. Also, the lower bound for the distributed case in
\eqref{eqn:lowerbndregret} initially increases and then decreases
with $U$ 
This is because as $U$ increases there are two competing effects:
decrease in regret due to decrease in number of $U$-worst channels
and increase in regret due to increase in number
of users visiting these $U$-worst channels.

Fig.\ref{fig:channels} evaluates the performance of the different
algorithms as the number of channels $C$ is varied while fixing the
number of users $U$. The probability of availability of each
additional channel is set higher than those already present. Here,
the regret monotonically increases with $C$ in all cases. When the
number of channels   increases along with the quality of the
channels, the regret increases as a result of an increase in the
number of $U$-worst channels as well as the increasing gap in
quality between the $U$-best and $U$-worst channels.

Also, the situation where the ratio $\frac{U}{C}$ is fixed to be
$0.5$ and both the number of users and channels along with their
quality   increase is considered in Fig.\ref{fig:upperbound}. As the
number of users  increases the regret   increases as the number of
channels $C$ and their quality are both increasing. Once again, this
is in agreement with theory as the number of $U$-worst channels
increases as $U$ and $C$ increase while keeping $\frac{U}{C}$ fixed.

\vspace{0.05in}
\subsubsection*{Collisions and Learning}

Fig.\ref{fig:known} verifies the logarithmic nature of the number
collisions  under the random allocation scheme $\randalloc$.
Additionally, we also plot the number of collisions under
$\randalloc$ in the ideal scenario when the channel availability
statistics $\mubf$ are known to see the effect of learning on the
number of collisions.  The low value of the  number of collisions
obtained under  known channel parameters in the simulations is in
agreement with theoretical predictions, analyzed as $U\Ebb[\Upsilon(U,U)]$ in
Lemma \ref{lemma:Pi}.   As the number of slots $n$ increases, the
gap between the number of collisions under the known and unknown
parameters increases since the former converges to a finite constant
while the latter grows as $O(\log n)$. The logarithmic behavior of
the cumulative number of collisions can be inferred from
Fig.\ref{fig:algocomp}. However, the curve in Fig.\ref{fig:known}
for the unknown parameter case  appears linear in $n$ due to the
small value of $n$.  \vspace{0.05in}

\subsubsection*{Difference between $g^{\mbox{\tiny OPT}}$ and
$g^{\mbox{\tiny MEAN}}$}

Since the statistic $g^{\mbox{\tiny MEAN}}$ used in the schemes in
this paper differs   from the optimal  statistic $g^{\mbox{\tiny
OPT}}$  in \eqref{eqn:gopt}, a simulation is done to compare the
performance of the schemes under both the statistics. As expected,
in   Fig.\ref{fig:statcomp}, the optimal scheme has better
performance. However, the use of   $g^{\mbox{\tiny MEAN}}$ enables
us to provide finite-time bounds, as described earlier.
\vspace{0.05in}

\begin{figure}[t] 
\bc\bp\psfrag{No. of  users}[c]{} \psfrag{Norm. 
Regret}[c]{}\psfrag{Normalized Regret (Rand)}[l]{\tiny Random 
Allocation Scheme}\psfrag{Normalized Regret (Cent)}[l]{\tiny Central 
Allocation Scheme}\psfrag{Normalized Regret (Pre)}[l]{\tiny 
Pre-Allocation Scheme}\psfrag{Theoretical Lower Bound 
(Distributed)}[l]{\tiny  Distributed Lower Bound}\psfrag{Theoretical 
Lower Bound (Centralized)}[l]{\tiny
 Centralized Lower Bound}\includegraphics[width=2.3in,height=1.5in]{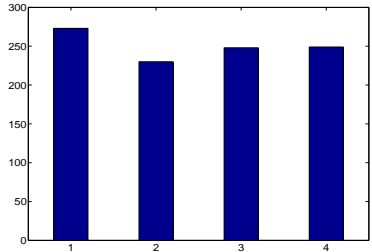}
\ep\ec
  \caption{Simulation Results. Probability of Availability $\mubf=[0.1,0.2,\ldots,
0.9]$. No. of slots where   user 
has best channel vs. user.  $U=4$, $C=9$, $n=2500$ slots, $1000$ runs, 
$\randalloc$.}  \label{fig:fairness}
 \end{figure}

\subsubsection*{Fairness}

One of the important features of $\randalloc$ is that it does not
favor any one user over another. Each user has an equal chance of 
settling down in any one of the  $U$-best channels. 
Fig.\ref{fig:fairness} evaluates the fairness characteristics of 
$\randalloc$. The simulation assumes $U = 4$ cognitive users vying 
for access to $C = 9$ channels. The graph depicts which user 
asymptotically gets the best channel over $1000$ runs of the random 
allocation scheme. As can be seen, each user has approximately the 
same frequency of being allotted the best channel indicating that the 
random allocation scheme is indeed fair.\vspace{0.05in}

\section{Conclusion}\label{sec:conclusion}

In this paper, we proposed novel  policies for distributed learning of channel availability statistics and  channel access of multiple secondary users in a cognitive network. The first policy assumed that the number of secondary users in the network is known, while the second policy removed this requirement.  We provide provable guarantees for our policies in terms of sum regret.  Combined with the lower bound  on regret for any uniformly-good learning and access policy, our first policy achieves order-optimal regret while our second policy is also nearly order optimal. Our analysis in this paper provides insights on incorporating learning and distributed medium access control in a practical cognitive network.

The results of this paper open up an interesting array of problems for future investigation.  Our assumptions of an i.i.d. model for primary user transmissions and   perfect sensing at the secondary users need to be relaxed. Our policy allows  for an  unknown but fixed number of secondary users, and it is of interest to incorporate  users dynamically entering and leaving the system.  Moreover, our model ignores dynamic traffic at the secondary nodes and extension to a queueing-theoretic formulation is desirable.
%
%
%

\subsection*{Acknowledgement}
The authors thank the guest editors and anonymous reviewers for valuable comments that vastly improved this paper, and for pointing out an error in Proposition~\ref{prop:bounds}. The authors thank  Keqin Liu and Prof. Qing Zhao for extensive discussions, feedback on the proofs in an earlier version of the manuscript, and for sharing their simulation code. The authors also thank Prof. Lang Tong and Prof. Robert Kleinberg at Cornell, Prof. Bhaskar Krishnamachari at USC and Dr. Ishai Menache at MIT for 
helpful comments.
\begin{appendix}

\subsection{Proof of
Theorem~\ref{thm:lowerbnd_cent}}\label{proof:lowerbnd_cent} The
result in \eqref{eqn:regret_cent_mean} involves extending the
results of \cite[Thm. 1]{Auer&etal:02ML}. Define $T_i(n)\defeq
\sum_{j=1}^U T_{i,j}(n)$ as  the number of times a channel $i$ is
sensed in $n$ rounds for all users. We will show that
\beq\label{eqn:NUCB} \Ebb[T_{i}(n)] \leq \sum_{k\in U
\mbox{\scriptsize-best}}\left[\frac{ 8 \log n}{\Delta(k^*,i)^2} +
1+\frac{\pi^2}{3}\right]  , \quad  \forall i\in U \mbox{-worst}.\eeq
  We have \begin{align}\nn &\Pbb[\mbox{Tx. in $i$ in $n^{\tha}$ slot}]
=\Pbb[g(U^*;n)\leq g(i;n)], \\ =& \Pbb[\Ac(i;n)\cap (g(U^*;n)\leq
g(i;n)) ] \nn \\ &+ \Pbb[\Ac^c(i;n)\cap (g(U^*;n)\leq g(i;n)) ],\nn
\end{align} where
\[\Ac(i;n)\defeq\bigcup_{k\in U \mbox{\scriptsize-best}}(g(k;n)\leq
g(i;n))\] is the event that at least one of the $U$-best channels
has $g$-statistic less than $i$. Hence, from union bound we have \[
\Pbb[\Ac(i;n)]\leq \sum_{k\in
U\mbox{\scriptsize-best}}\Pbb[g(k;n)\leq g(i;n)].\] We have for $C>
U$,
\[\Pbb[\Ac^c(i;n)\cap
(g( U^*;n)\leq g(i;n)) ]=0,\] Hence, \[ \Pbb[\mbox{Tx. in $i$ in
$n^{\tha}$ round}] \leq \sum_{k\in
\mbox{\scriptsize-best}}\Pbb[g(k;n)\leq g(i;n)].\] On  the lines of
\cite[Thm. 1]{Auer&etal:02ML}, we have $\forall k,i:k \mbox{ is
$U$-best}, i \mbox{ is $U$-worst}$ \[ \sum_{l=1}^n I[g(k;l)\leq
g(i;l)] \leq   \frac{ 8 \log n}{\Delta(k^*,i)^2} + 1+\frac{\pi^2}{3}
.\] Hence, we have  (\ref{eqn:NUCB}). For the bound on regret, we
can break $R$ in \eqref{eqn:regret_def} into two terms
\begin{align}\nn R(n;\mubf,U, \cent) =& \sum_{i\in U\mbox{\scriptsize-worst}}
\Bigl[ \frac{1}{U}\sum_{l=1}^U \Delta(l^*, i)\Bigr] \Ebb[T_i(n)]\\
&+ \sum_{i\in U \mbox{\scriptsize-best}} \Bigl[
\frac{1}{U}\sum_{l=1}^U \Delta(l^*,i)\Bigr] \Ebb[T_i(n)]. \nn
\end{align} For the second term, we have \begin{align} \nn &
\sum_{i\in U\mbox{\scriptsize-best}} \Bigl[ \frac{1}{U}\sum_{l=1}^U
\Delta(l^*,i)\Bigr]  \Ebb[T_i(n)]\\ &\leq  \Ebb[T^*(n)] \sum_{i\in U
\mbox{\scriptsize-best}} \Bigl[ \frac{1}{U}\sum_{l=1}^U \Delta(l^*,
i)\Bigr]  = 0,\nn\end{align}   where $T^*(n) \defeq
\max\limits_{i\in U \mbox{\scriptsize-best}} T_i(n)$. Hence, we have
the bound.\qed

\subsection{Proof of Proposition~\ref{prop:bounds}}\label{proof:bounds}
For convenience, let $T_i(n) := \sum_{j=1}^U T_{i,j}(n)$, $V_i(n) := \sum_{j=1}^U V_{i,j}(n)$.   Note that $ \sum_{i=1}^C T_{i}(n) = nU,$ since each user selects one channel for sensing in each slot and there are $U$ users. 
From \eqref{eqn:regret_collision},\begin{align}R(n) =& n\sum_{i=1}^U \mu(i^*) -\sum_{i=1}^C \mu(i)\Ebb[V_i(n)],\nn\\ \leq& \sum_{i\in U\mbox{\scriptsize-best}} \mu(i)(n - \Ebb[V_i(n)]) \nn \\ \label{eqn:1} \leq & \mu(1^*) (nU - \sum_{i\in U\mbox{\scriptsize-best}} \Ebb[V_i(n)])\\ =&\mu(1^*)( \Ebb[M(n)] + \sum_{i\in U\mbox{\scriptsize-worst}}\Ebb[T_i(n)]) \label{eqn:2},
\end{align}where Eqn.\eqref{eqn:1} uses the fact that $V_i(n) \leq n$ since total number of sole occupancies in $n$ slots of channel $i$ is at most $n$, and Eqn.\eqref{eqn:2} uses the fact that $M(n) = \sum_{i\in U\mbox{\scriptsize-best}}(T_i(n) - V_i(n))$.
     
For the lower bound, since each user selects one channel for sensing
in each slot, $\sum_{i=1}^C \sum_{j=1}^U T_{i,j}(n) = nU$. Now
$T_{i,j}(n) \geq V_{i,j}(n)$.  \begin{align}\nn
R(n;\mubf,U,\policy)\geq& \frac{1}{U} \left[\sum_{k =1}^U
\sum_{j=1}^U \sum_{i=1}^C \Delta(U^*,i) \Ebb[T_{i,j}(n)] \right],\\
\geq& \sum_{j=1}^U \sum_{i\in
U\mbox{\scriptsize-worst}}\Delta(U^*,i)
\Ebb[T_{i,j}(n)].\nn\end{align}\qed

\subsection{Proof of Lemma \ref{lemma:Pi}}\label{proof:Pi}

Although, we could directly compute the time to absorption of the Markov chain, we give a simple bound $\Ebb[\Upsilon(U,U)]$ by considering an i.i.d process over the same state space. 
We  term this process as  a genie-aided modification of random allocation
scheme, since this can be realized as follows: in each slot, a genie  checks if any collision
occurred, in which case, a new random variable is drawn from
$\unif(U)$ by all users. This is in contrast to the original random allocation scheme
where a new random variable is drawn only when the particular user
experiences a collision. Note that for $U=2$ users, the two
scenarios coincide.

For the genie-aided scheme, the expected number of slots to hit
orthogonality is just the mean of the geometric distribution  \beq
\sum_{k=1}^\infty k (1-p)^{k} p =\frac{1-p}{
p}<\infty,\label{eqn:geomean}\eeq where $p$ is the probability of
having an orthogonal configuration in a slot. This is in fact the
reciprocal of the number of {\em compositions} of $U$  \cite[Thm.
5.1]{Bona:book}, given by \beq \label{eqn:p} p =  \binom{
2U-1}{U}^{-1}.\eeq The above expression is nothing but the
reciprocal of number of ways $U$ identical balls (users)
can be placed in $U$ different bins (channels): there are $2U-1$
possible positions to form $U$ partitions of the balls.

Now for the random allocation scheme without the genie, any user not
experiencing collision does {\em not} draw a new variable from
$\unif(U)$. Hence, the number of possible configurations in any slot
is lower than under genie-aided scheme. Since there is only one
configuration satisfying orthogonality\footnote{since all users are
identical for this analysis.}, the probability of orthogonality
increases in the absence of the genie and is at least \eqref{eqn:p}.
Hence, the number of slots to reach orthogonality without the genie
is at most \eqref{eqn:geomean}. Since in any slot, at most $U$
collisions occur, \eqref{eqn:Pi} holds.\qed

\subsection{Proof of Lemma~\ref{lemma:numintfrandalloc}}\label{proof:numintfrandalloc}

Let $c_{n,m} \defeq \sqrt{\frac{2 \log n}{m}}$.

{\bf Case 1: } Consider $U=C=2$ first. Let \[ \Ac(t,l)\defeq
\{g_j^{\mbox{\tiny MEAN}}(1^*;t-1)\leq g_j^{\mbox{\tiny
MEAN}}(2^*;t-1), T'_{j}(t-1) \geq l\}.\] On lines of \cite[Thm.
1]{Auer&etal:02ML},  {\begin{align} T'(n) &\leq    l +
\sum_{t=2}^n   I[ \Ac(t,l)],\nn \\   & \leq l+
\sum_{t=1}^\infty\sum_{m+h=l}^{t}    I\left( \bar{X}_{1^*,j}(h) +
c_{t,h}  \leq  \bar{X}_{2^*,j}(m) + c_{t,m}\right).
\nn
\end{align} } The above event is implied by  \[
\bar{X}_{1^*,j}(h)+c_{t,h}\leq \bar{X}_{2^*,j}(h)+c_{t,h+m}\] since
$c_{t,m}>c_{t,h+m}$.

The above event implies at least one of the following events and
hence, we can use the union bound. \begin{align}\nn
\bar{X}_{1^*,j}(h)&\leq \mu_{1^*}-  c_{t,h}  ,\nn\\
\bar{X}_{2^*,j}(m)&\geq \mu_{2^*} + c_{t,h+m},\nn\\ \nn
\mu_{1^*}&<\mu_{2^*} +  2c_{t,h+m}.\end{align} From
the Chernoff-Hoeffding bound,
\begin{align}\Pbb[\bar{X}_{1^*,j}(t)\leq \mu_{1^*}- c_{t,h}]&\leq
t^{-4},\nn \\ \Pbb[\bar{X}_{2^*,j}\geq \mu_{2^*} + c_{t,h+m}]&\leq
t^{-4},\nn
\end{align} and    the event that
 $ \mu_{1^*}<\mu_{2^*} + 2c_{t,h+m}$ implies that \beq \nn h+m <
\left\lceil\frac{8 \log t}{\Delta_{1^*,2^*}^2} \right\rceil.  \eeq
Since
\[\sum_{t=1}^\infty \sum_{m=1}^{t} \sum_{h=1}^{t}
2t^{-4} = \frac{\pi^2}{3},\],
 \[ \Ebb[T'(n; U=C=2)]\leq \frac{8 \log
n}{\Delta_{1^*,2^*}^2}+1 + \frac{\pi^2}{3}. \]

{\bf Case 2: } For $\min(U,C)>2$, we have \[ T'(n) \leq U\sum_{a=1}^U \sum_{b=a+1}^C \sum_{m=1}^n  I(g_j^{\mbox{\tiny MEAN}}(a^*;m)<
g_j^{\mbox{\tiny MEAN}}(b^*;m)),\]where $a^*$ and $b^*$ represent
channels with $a^{\tha}$ and $b^{\tha}$ highest availabilities. On
lines of the result for $U=C=2$, we can show that
\[ \sum_{m=1}^n \Ebb I[g_j^{\mbox{\tiny MEAN}}(a^*;m)<
g_j^{\mbox{\tiny MEAN}}(b^*;m)]\leq
 \frac{8 \log
n}{\Delta_{a^*,b^*}^2}+1 + \frac{\pi^2}{3}.\]  
Hence, \eqref{eqn:tdashj} holds.\qed

\subsection{Proof of Theorem \ref{thm:intfrandalloc}}\label{proof:thmintfrandalloc}
Define the good event as all users having correct top-$U$ order of the $g$-statistics, given by
\[ \Gmsc(n)
\defeq \bigcap_{j =1}^U \{\mbox{Top-$U$ entries of }\bfg_j(n) \mbox{
are same as in }\mubf\}.\]The number of slots under the  bad event is\[ \sum_{m=1}^n I[\Gmsc^c(m)] = T'(n),\] by definition of $T'(n)$. In each slot, either a good or a bad event occurs. Let $\gamma$ be
the total number of collisions in $U$-best channels between two bad
events, \ie under a run of good events.  In this case,   all the  users
have the correct top-$U$ ranks of channels and hence,\[
\Ebb[\gamma|\Gmsc(n)]\leq U \Ebb[\Upsilon(U,U)]<\infty,\]where $\Ebb[\Upsilon(U,U)]$ is given by
(\ref{eqn:Pi}). Hence, each transition from the bad to the good state
results in at most $U \Ebb[\Upsilon(U,U)]$ expected number of collisions in the $U$-best
channels. The expected number of collisions under the bad event is at most $U \Ebb[T'(n)]$. Hence,    \eqref{eqn:M} holds.\qed

\subsection{Proof of
Lemma~\ref{lemma:cond_regret}}\label{proof:cond_regret}

Under $\Cc(n;U)$, a $U$-worst channel is sensed only if it is mistaken to be a $U$-best channel.
Hence, on lines of
Lemma~\ref{lemma:uworstrandalloc},\[
\Ebb[T_{i,j}(n)|\Cc(n;U)]  =O(\log n), \quad \forall i \in U\mbox{-worst}, j=1,\ldots, U.\] 
For the number of collisions $M(n)$ in the $U$-best channels,  there can be at most $U \sum_{k=1}^a\xi(n;k)$  collisions in the $U$-best channels where $a:= \max_{j=1,\ldots, U} \widehat{U}_j$ is the maximum estimate of number of users. Conditioned on $\Cc(n;U,)$, $a\leq U$, and hence, we have \eqref{eqn:cond_m}.\qed
 
\subsection{Proof of Proposition~\ref{prop:numcoll}}\label{proof:numcoll}

Define the good event as all users having correct top-$U$ order, given by
\[ \Gmsc(n)
\defeq \bigcap_{j =1}^U \{\mbox{Top-$U$ entries of }\bfg_j(n) \mbox{
are same as in }\mubf\}.\]The number of slots under the  bad event is\[ \sum_{m=1}^n I[\Gmsc^c(m)] = T'(n),\] by definition of $T'(n)$. In each slot, either a good or a bad event occurs. Let $\gamma$ be
the total number of collisions in $k$-best channels between two bad
events, \ie under a run of good events.  In this case,   all the  users
have the correct top-$U$ ranks of channels and hence,\[
\gamma|\Gmsc(n)\leqst U \Upsilon(U,k),\]   The number of collisions under the bad event is at most $T'(n)$. Hence,    \eqref{eqn:numcoll} holds.\qed
 
\subsection{Proof of Lemma~\ref{lemma:underestimate}}\label{proof:underestimate}

We are interested in \begin{align}\nn &\Pbb[\Cc^c(n);U] = \Pbb[\cup_{j=1}^U \widehat{U}^\est_j(n) > U],\nn\\ &= \Pbb[\bigcup_{m=1}^n\bigcup_{j=1}^U \{\Phi_{U,j}(m)> \xi(n;U)\}] \nn,\\ &=   \Pbb[\max_{j=1,\ldots,U}\Phi_{U,j}(n)> \xi(n;U) ],\nn \end{align} where $\Phi$ is given by \eqref{eqn:Phi}. For $U=1$, we have $\Pbb[\Cc^c(n);U]=0$ since no collisions occur.

Using \eqref{eqn:numcoll} in Proposition~\ref{prop:numcoll}, 
\begin{align}&\Pbb[\max_{j=1}^k\Phi_{k,j}(n)> \xi(n;k) ]\nn\\&\leq\Pbb[k\Upsilon(U,k)(T'(n)+1)>\xi(n;k)]\nn\\&\leq \Pbb[k(T'(n)+1)> \frac{\xi(n;k)}{\alpha_n}] + \Pbb[\Upsilon(U,k)>\alpha_n]\nn \\ &\leq \frac{ k\alpha_n(\Ebb[T'(n)]+1)}{\xi(n;k)}+ \Pbb[\Upsilon(U,k)>\alpha_n], \label{eqn:3}\end{align}
using Markov inequality. By choosing $\alpha_n = \omega(1)$,  the second term in \eqref{eqn:3}, viz., $\Pbb[\Upsilon(U,k)>\alpha_n] \to 0$ as $n \to \infty$, for $k\geq U$. For the first term, from \eqref{eqn:Tdashdash} in Lemma~\ref{lemma:Tdashdash}, $\Ebb[T'(n)] = O(\log n)$. Hence, by choosing $\alpha_n = o(\xi^*(n;k)/\log n)$, the first term decays to zero.  Since $\xi^*(n;U) = \omega(\log n)$, we can choose $\alpha_n$ satisfying both the conditions. By letting $k=U$ in \eqref{eqn:3}, we have $\Pbb[\Cc^c(n);U] \to 0$ as $n \to \infty$, and
\eqref{eqn:underestimate} holds. \qed

\end{appendix}


\end{document}